\documentstyle[12pt,aaspp4]{article}

\newcommand{\ts}{\thinspace}

\newcommand{\p}{$^{\prime}$\ }
\newcommand{\nul}{$\nu${\it L}$_{\nu}$}
\newcommand{\msun} {\hbox{M$_\odot$}\ }
\newcommand{\msunns}{\hbox{M$_\odot$}}

\newcommand{\lsun}{\hbox{{\it L}$_\odot$}\ }
\newcommand{\lsunns}{\hbox{{\it L}$_\odot$}}
\newcommand{\rqu}{r$^{-1/4}$\ }
\newcommand{\sersic}{S\'{e}rsic }

\lefthead{Surace et al.}
\righthead{Imaging of ULIGs in the near-UV }

\begin{document}

\title{Imaging of Ultraluminous Infrared Galaxies in the Near-UV}

\author{Jason A. Surace}
\affil{Infrared Processing and Analysis Center, MS 100--22, California Institute of 
Technology, Jet Propulsion Laboratory, Pasadena, CA 91125 \\
University of Hawaii, Institute for Astronomy, 2680 Woodlawn Dr., 
Honolulu, HI, 96822 \\
Electronic mail: jason@ipac.caltech.edu
and \\}
\author{D. B. Sanders}
\affil{University of Hawaii, Institute for Astronomy, 2680 Woodlawn Dr., 
Honolulu, HI, 96822 \\
Electronic mail: sanders@ifa.hawaii.edu}

\ \ 

\centerline{{To appear in the August, 2000 edition of the {\it Astronomical Journal}}}

\begin{abstract}

We present the first ground-based U\p ($\lambda_{central}=$3410\AA, 
$\Delta_{\lambda}$=320\AA) images of Ultraluminous Infrared Galaxies (ULIGs). Two 
samples were observed: (1) ``warm'' ULIGs with mid-infrared colors characteristic of 
active galactic nuclei ({\it f}$_{\rm 25\mu m}$/{\it f}$_{\rm 60\mu m} > $0.2), which 
are believed to represent a critical transition phase between cooler ULIGs and optically 
selected QSOs according to a previously proposed evolutionary model, and (2) the 
complementary ``cool'' ULIGs ({\it f}$_{\rm 25\mu m}$/{\it f}$_{\rm 60\mu m} < $0.2) 
which in the evolutionary model are the progenitors of warm ULIGs and which have 
many characteristics associated with active star-formation. Although in some cases 
there is also emission identified with an active galactic nucleus, the U\p emission 
originates primarily in massive young stars and as such allows a direct examination of 
the sites of recent high-mass star formation. Strong U\p emission (median total {\it 
M}$_{\rm U}$ = -20.8) is seen in all systems and in some cases the extended tidal 
features (both the smooth stellar distribution and compact star-forming features) 
contribute up to 60--80\% of the total flux. The star-forming regions in both samples 
are found to have ages based on spectral synthesis models in the range 10-100 Myrs, 
and most differences in color between them can be attributed to the effects of dust 
reddening. Additionally, it is found that star-formation in compact knots in the tidal 
tails is most prominent in those ULIGs which have double nuclei, suggesting that the 
star-formation rate in the tails peaks prior to the actual coalescence of the galaxy 
nuclei and diminishes quickly thereafter. Similar to results at other wavelengths, the 
observed star formation at U\p can only account for a small fraction of the known 
bolometric luminosity of the ULIGs. Azimuthally averaged radial light profiles at U\p 
are characterized by a \sersic law with index {\it n}=2, which is intermediate between 
an exponential disk and an \rqu law and closely resembles the latter at large radii. The 
implications of this near-ultraviolet imaging for optical/near-infrared observations of 
high redshift counterparts of ULIGs are discussed.

\end{abstract}


\section{Introduction}

Many recent Hubble Space Telescope ({\it HST}) observations have shown the 
prevalence of clustered star-formation in interacting galaxies (Whitmore et al. 1995, 
Meurer et al. 1995). Other observations have shown that Ultraluminous Infrared 
Galaxies (ULIGs; objects with infrared luminosities, $L_{\rm ir}$ 
\footnote{$L_{\rm ir} \equiv$ {\it L}(8--1000\micron)
is computed using the flux in all four {\it IRAS} bands according to the
prescription given in Perault \markcite{perault} (1987); see also Sanders \& Mirabel 
(1996).  Throughout this paper we use $H_{\rm o}$ = 75 km s$^{-1}$Mpc$^{-1}$, q$_{\rm 
o}$ = 0.5 (unless
otherwise noted).}, greater than
$10^{12}\ L_{\sun}$) often have extended morphologies consistent with being 
advanced mergers (Sanders et al. 1988a, Murphy et al. 1996, Clements et al. 
1996). Surace et al. (1998, hereafter Paper I) showed that these same systems 
have evidence for ``knots'' of star-formation with characteristic diameters of 
$\approx$100 pc distributed over their central several kpc and along their 
extended tidal features. Estimated ages for these knots based on optical and near-
infrared colors were in the range of 10--800 Myrs. In at least 25\% of ULIGs, 
particularly those with ``warm'' mid-infrared colors and Seyfert optical spectra, it is 
also possible that some fraction of the optical and hence U\p light is emitted by an AGN.

With the exception of integrated photometry taken with single beam photometers for a 
few systems (Neugebauer et al. 1987, Young et al. \markcite{young} 1996), only two of 
the ULIGs have been examined at short wavelengths before (Trentham \markcite{neil}
et al. 1999). This is primarily due to former assumptions about the dusty nature of the 
ULIGs. Since the bulk of the luminosity in ULIGs is emitted at 25--100 $\mu$m as a 
result of dust absorption and reradiation, the assumption has always been that the 
galaxies themselves are intrinsically dusty and that extinction must therefore be very 
high. In short, there would be nothing to see at short wavelengths. However, ground-
based and {\it HST} imaging (Paper I, Surace 1998, Surace \& Sanders 1999; hereafter 
Paper II) 
has revealed the presence of luminous clustered star formation at optical and 
near-infrared wavelengths. Paper II developed a broad-band multi-color 
technique at optical and near-infrared wavelengths in order to measure and 
correct for the presence of a foreground dust screen. These observations 
have shown that the apparent line-of-sight extinction is quite low (typically {\it 
A}$_{\rm V} <$ 2 magnitudes), implying that many of the active star forming 
regions have already ejected any surrounding dust or have particularly 
favorable scattering geometries. Additionally, millimeter-wave observations of 
some of these systems have indicated directly that the dust is strongly 
centrally concentrated and localized, and hence the more extended starburst 
regions may be relatively unobscured (Bryant \& Scoville 1996, Scoville et al. 
\markcite{scoville} 
1997). Line-of-sight extinction is therefore not a serious impediment to 
detecting the ULIG star-forming regions at short wavelengths.

There are several advantages to working at shorter wavelengths. As noted in 
Paper I, there is an ambiguity in interpreting broad-band colors of starbursts 
with burst ages of 10--100 Myrs, as their colors in all filters redder than {\it B} 
are degenerate during this time period. This can be partially disentangled through the 
use of U-band imaging; ({\it U$-$B}) colors evolve almost monotonically 
throughout this period in the life of the starburst. Although reddening will 
still prevent a precise starburst age estimate since it is most severe at short 
wavelengths, the ({\it U$-$B}) colors, in conjunction with longer wavelength 
data ( Surace et al. 1998 [Paper I], Surace et al. 1999 [Paper II]), will help constrain the 
upper age limits 
of the starbursts more accurately. Furthermore, the youngest stars, which 
dominate the luminosity of the starburst, have spectral energy distributions 
(SEDs) which peak at ultraviolet wavelengths. Ultraviolet observations 
therefore provide the greatest possible contrast between compact young star-
forming regions and the redder, extended underlying galaxy. In the evolutionary 
model of Sanders et al. (1988a), warm ULIGs are a more advanced evolutionary state 
than cool ULIGs, and this should be reflected in the upper limits of the knot ages.

Finally, there have been many recent advances in the study of high redshift galaxies. 
Particularly important are deep submillimeter surveys, which have discovered a 
significant population of high-redshift galaxies which are luminous at far-infrared 
wavelengths (Smail et al. 1997, Hughes et al. 1998), and which are inferred on the basis 
of their submillimeter luminosity to be high-redshift counterparts of the local ULIGs. 
Optical and near-infrared observations of these high-redshift objects correspond to 
rest-frame UV, and hence a better understanding of the UV properties of local ULIGs is 
required in order to interpret high-redshift results.

We present U\p imaging data for selected ULIGs drawn from two complete 
samples of ``warm'' AGN-like ULIGs and ``cool''  starburst-like ULIGs which 
have been previously well-studied at other wavelengths (Sanders et al. 1988ab, 
Kim et al. 1998, Paper I, II, Surace 1998), with the primary goal of characterizing these 
systems at near-UV wavelengths in order to further understanding of high-redshift 
systems. U\p and optical colors are used to derive 
starburst knot ages assuming synthetic starburst models. The estimated ages, 
reddenings, and luminosities are used to derive the contribution of the 
extended starburst regions to the known bolometric luminosities of the galaxy 
systems. The history of star-formation in these systems and the implications of these 
new observations for high-redshift systems are 
discussed. 

\section{Data}

\subsection{Sample Selection}

The ultraluminous infrared galaxies were chosen from two small, nearly 
complete samples. The first sample consists of the ``warm'' ULIGs first 
described by Sanders et al. (1988b); more detailed optical and near-infrared 
imaging can be found in Surace et al. (1998) and Surace \& 
Sanders (1999). These systems were selected using a mid-infrared color 
criterion known to select active galaxies ($f_{25}/f_{60} > 0.2$)
\footnote{The quantities $f_{12}$,
$f_{25}$, $f_{60}$, and $f_{100}$ represent the {\it IRAS} flux densities in Jy
at 12{\ts}\micron, 25{\ts}\micron, 60{\ts}\micron, and 100{\ts}\micron\
respectively.}
and which has been used previously to select AGN from the IRAS Point 
Source Catalog (de Grijp et al. 1985, 1987). Studies of several small but complete 
samples of warm ULIGs have shown that many of
these objects have a point-like optical appearance on the Palomar Sky Survey
and that they exhibit broad (i.e. Seyfert 1) optical emission lines, 
characteristics that have led them to be referred to as ``infrared QSOs'' (e.g. 
Low et al. \markcite{low} 1988, Sanders et al. 1988b). These warm objects, which
account for $\sim${\ts}20--25{\ts}\% of the total population of ULIGs discovered 
by {\it IRAS} (Kim 1995), appear to represent a critical transition stage in the 
evolution of the larger population of ``cool'' ULIGs into optical QSOs according 
to the unification scheme proposed by Sanders et al. (1988a). Since they were 
 selected to be transition objects similar to optical QSOs, the luminosity criterion used 
was 
{\it L}$_{\rm bol} > 10^{12}$ \lsun, hence the inclusion of two 
systems with {\it L}$_{\rm ir} < 10^{12}$ \lsunns (Sanders et al. 1988b).

The second sample consists of ULIGs with complementary ``cool'' colors 
($f_{25}/f_{60} < 0.2$) as detailed by Surace et al. (2000). These systems are more 
similar to the majority of ULIGs (representing roughly 75\% of the ULIG 
population; Kim 1995) and are thus similar to the Bright Galaxy Sample ULIGs 
described by Sanders et al. (1988a). They typically have HII or LINER 
spectra, have diffuse nuclear regions with optical/near-infrared colors 
similar to young stars obscured by {\it A}$_{\rm V} \approx$1-4 magnitudes 
(Surace 1998), and a high percentage of double nuclei. They therefore appear 
to represent an earlier stage in the merger process. Together these two 
samples span the entire range of interactions and far-infrared colors 
observed in ULIGs and in the evolutionary scheme of Sanders et al. (1988ab) represent 
two temporally distinct populations of the same type of physical object.
Since the ultraviolet emission originates predominantly in massive young stars which 
have formed on timescales similar to that of the merger process, if the evolutionary 
hypothesis is correct then an examination of the warm and cool ULIGs should provide 
evidence for differences in their ultraviolet properties.

Because of spatial constraints and the limited telescope resources available, not all of 
the objects in 
the two complete samples could be observed. A total of 
20 objects were observed: 11 (of a complete sample of 12) ``warm'' ULIGs and 9 (of a 
sample of 14) cool ULIGs. Tables 1 \& 2 list the objects actually observed in the two 
samples, along with observation details.

\subsection{Observations}

The data were taken on December 26--29, 1997, March 23--24, 1998 and 
November 12--13, 1998 at the f/31 focus of the UH 2.2m telescope on Mauna Kea 
using the Orbit 2048$\times$2048 camera. The Orbit camera is a backside-
illuminated, thinned CCD that is UV-flooded for normal operation; it's quantum 
efficiency at U\p (based on laboratory tests) is as high as 80\%. At the f/31 
focus this camera is oversampled, and hence the data was binned 2$\times$2 
on chip yielding a pixel scale of 0.09\arcsec \ pixel$^{-1}$. The UH U\p filter 
(which is noticeably bluer than the standard Johnson U filter; 
$\lambda_{central}=$3410\AA, $\Delta_{\lambda}$=320\AA ;Wainscoat 
\markcite{wainscoat} 1996) was used for all 
observations.
Total exposure times were typically 30--90 minutes, the long exposure times 
being necessitated by the narrowness of the U\p filter. A sequence of five or 
more frames was taken in order to allow post-processing rejection of the high 
cosmic ray incidence. These frames were dithered to further decorrelate 
structured pixel-to-pixel noise in the CCD itself (such as blocked rows, flat-
fielding errors, etc.). The CCD bias pattern 
was removed by subtracting a median bias frame constructed from 20--30 bias 
frames (zero-time exposures), and pixel-to-pixel response variations were corrected by 
dividing the data by a high S/N median twilight flat. In both cases the expected poisson 
noise in the 
calibration frames was below 0.5\%.
On the second run this camera developed severe dark current instabilities. 
This was
calibrated by taking 4--5 hours of dark current (closed shutter integration) 
data at the beginning and end of each night. The high 
S/N dark current images were made for each half of the night, scaled to the 
exposure times used for the science data, and subtracted from them. In most 
cases these dark integrations had the same exposure times as used for the 
actual science observations, thus helping eliminate time-dependent dark 
current variations. This strategy was effective with no visible 
dark current pattern after processing. The individual frames were then 
rotated to the normal orientation (northeast at upper left), aligned using 
IMALIGN in IRAF, and medianed using IMCOMBINE and an algorithm that 
rejects cosmic rays based on the CCD noise characteristics.

The data were flux calibrated by observing optical standard stars (Landolt 
\markcite{landolt83} 1983, \markcite{landolt92} 
1992) with a wide range of colors and at different airmasses, allowing the 
derivation of accurate zero-point offsets and color terms relating the U\p 
filter to the standard Johnson U-band filter. This also allowed the atmospheric 
extinction to be measured and corrected; this can be as high as 0.2--0.6 
magnitudes per airmass (the observed mean was 0.27) at such short 
wavelengths. In general, the photometric calibration appears to be accurate to 
0.04 magnitudes. Possibly the largest uncertainty is the conversion from U\p 
to Johnson U-band; the relatively large color term can introduce uncertainties 
as high as 0.05--0.1 magnitudes, depending on the underlying SED of the 
observed feature. Throughout this paper quoted magnitudes will be on the 
Johnson U-band calibration, using the color correction :

\begin{equation}
\centerline{m$_U$ = -2.5 Log$\left( {adu}\over{exptime}\right)$ + 
ZP$_{U^{\prime}}$ - (airmass$\cdot$extinction$_{U^{\prime}}$) + 
0.13$\cdot$({\it B-I})} 
\end{equation}

\noindent where the first term is the instrumental magnitude, the second is the 
instrumental magnitude zeropoint, the third is the atmospheric extinction, and the last 
term is the color dependency which is appropriate for the combination of the Orbit 
camera and 
the UH U\p filter. No K-corrections have been applied to any magnitude 
reported here since the unknown SED at U\p prevents an 
accurate determination of such corrections. (B$-$I) is derived from previous 
optical observations (Surace et al. 1998, Surace 1998) and is accurately known (with 
values of (B$-$I) typically just 
greater than unity). For the integrated galaxy magnitudes we used the 
integrated galaxy colors for the color term correction, while for the star-forming 
regions we used the local 
colors. 

Fluxes were measured using aperture photometry. Circular and polygonal 
apertures were constructed using SKYVIEW. The background was estimated 
using a median of the data directly outside the apertures, excluding any bright 
small-scale structure. Integrated magnitudes were derived from large aperture 
measurements. In all cases the galaxies have been deeply imaged at I-band (Surace 
1998). The apertures were chosen to be large enough to encompass the entire known 
extent of the galaxies. The knots were identified based on spatial coincidence with 
previously identified star-forming knots seen in optical and near-infrared images. In 
many cases the spatial resolution is too poor to resolve all of the known knots. In these 
cases, apertures were chosen to be large enough to encompass the entire extent of each 
group of unresolved knots (and whose combined photometry are listed with a ``/'' in 
Table 3). Aperture corrections were derived from the observed 
point spread function (PSF) and have been applied to the magnitudes reported 
here.

Calibration of the PSF is problematic at U\p because most stars are not very 
luminous at such short wavelengths, thereby decreasing the likelihood of 
achieving a PSF S/N comparable to that of the target objects. In some cases it 
was possible to estimate the PSF directly from the combined science images in 
the same manner as in Paper II. DAOPHOT was used to iteratively construct an 
empirical PSF based on all the observed stars in the combined data, resulting 
in a high S/N PSF. In the remaining cases where not enough stars were 
present the PSF was derived from the standard star observations taken 
immediately before and after each set of science data. Most of the image 
degradation at short wavelengths is due to high-order atmospheric 
turbulence, and hence the quality of the off-axis tip/tilt guiding is largely 
irrelevant to the achieved PSF; the observed flux standards should therefore 
provide adequate PSF calibrators. Unfortunately, because of the poor 
resolution at short wavelengths, aperture corrections tend to be quite large 
(0.5--1 mag) for sub-arcsecond apertures, and hence knowledge of the PSF is 
critical. In most cases, in order to derive accurate colors the previous B and I-
band imaging had to be re-evaluated in apertures larger than those previously 
used in Surace (1998) in order to compensate for seeing-induced aperture 
effects. This was particularly necessary for comparison with the WFPC2 data. 
Total cumulative uncertainties, including those in calibration, measurement, 
and aperture corrections, are $\approx$0.2 magnitudes.

\section{Results}

\subsection{Morphology}

Figures 1 \& 2 present the U\p observations of the 11 warm and 9 cool ULIGs, 
respectively, and Appendix A provides detailed descriptions.
Many of the systems show considerable structure in their extended tidal 
features. In general, the morphology of the ULIGs at U\p is very similar to 
their morphology at B and to a lesser extent at I-band (Paper I). This is not 
surprising, as the extinction at U\p ({\it A}$_{\rm U}$=1.53{\it A}$_{\rm V}$) 
is only slightly larger than that at B ({\it A}$_{\rm B}$=1.45 {\it A}$_{\rm V}$), 
yet is much larger than I ({\it A}$_{\rm I}$=0.48 {\it A}$_{\rm V}$; Rieke \& 
Lebofsky \markcite{rieke} 1985). Therefore, dust extinction effects are unlikely to 
significantly 
alter the relative morphology between U\p and B. Any features near the I-band 
detection limit, however, will be completely obscured at U\p by even just a few 
magnitudes of extinction. Because of the generally poor 
spatial resolution at U\p (0.7--1.5\arcsec), any features within a radius of 
$\approx$1.5-2 kpc of the luminous point-like nuclei of some of the warm 
ULIGs will be undetectable.

Of the ``warm'' ULIGs, the two double nucleus systems Mrk{\ts}463 (3.8 kpc 
nuclear separation) and IRAS{\ts}08572+3915 (6.2 kpc nuclear separation) 
have the most spectacular U\p  features. In both cases up to half a dozen U\p 
knots are scattered along the lengths of the tidal tails. The extra-nuclear (diameter 
greater than 2.5 kpc)
regions account for 65\% and 83\% of the total U\p flux, respectively, where this 
number includes both the compact knots and the extended tidal emission. Similarly, the 
single active nucleus galaxies
I Zw 1, Mrk 1014 and Mrk 231 have U\p knots embedded in apparent tidal features. IRAS 
12071$-
$0444 and IRAS 15206+3342 have evidence for marginally resolved knots in 
their inner 10 kpc diameter nuclear regions. 
IRAS 05189$-$2524, Pks 1345+12, and IRAS 01003$-$2238 have no detectable knots, 
although this may be due to our limited spatial resolution.
The one warm ULIG which shows no U\p knots at all 
and should, based on the U\p-band  spatial resolution and detection limits, is 
IRAS 07598+6508. In 7 out of 11 cases more than 50\% of the U\p flux originates within 
a radius of 4 kpc of the nucleus, which more than anything reflects that 4 of the 
sample are acknowledged to be QSOs, with UV-luminous active nuclei.

The ``cool'' ULIGs have an equally varied U\p morphology. More than half of the cool 
ULIGs have more than 50\% of their U\p-band flux emitted in their inner 4 kpc radii. 
The single nucleus systems IRAS 00091$-$0738, UGC 5101, Mrk 273, and IRAS 
23365+3604 have long, well-developed tidal tails which are easily at I-band (Surace et al. 
2000), yet none of these tails has the 
high surface brightness U\p knots found in the warm systems IRAS 08572+3915 and 
Mrk 
463 and even the extended tidal structure itself is barely detected at U\p. The double 
nucleus systems IRAS 12112+0305 (2.5 kpc nuclear separation) 
and IRAS 14348$-$1447 (5.5 kpc nuclear separation), however, clearly 
have U\p knots located in their tails and also in their nuclear regions. This indicates 
that the appearance of knots in the tidal structure is more a function of dynamical age 
as identified by the appearance of double nuclei than it is of mid-IR color. Finally, 
the very widest separation system examined, IRAS 01199$-$2307 (24 kpc), has 
no evidence for U\p knots in its barely detectable tails.

This dichotomy in the appearance or lack of knots along the extended tidal features 
may 
indicate differences in the star-formation rate in the tails during different 
phases of the merger evolution. 
Since every merger system must pass through a double-nucleus phase prior to when 
the two nuclei have coalesced into a single nucleus, the double nucleus systems are 
dynamically younger than the single-nucleus systems. However, knots embedded in the 
extended tidal tail structure are visible in 5 out 7 of the double nucleus systems, but 
only 2 out of 13 of the single nucleus systems (Table 4). Based on Monte Carlo 
simulations, if the probability of having extranuclear knots were the same for both 
samples, the likelihood of our finding as few single nucleus systems and as many 
double nucleus systems with knots is less than 1\%, regardless of the lifetime of the 
knots vs. the merger lifetime.  This appearance of compact star-forming knots in the 
tails of every 
merger system with nuclear separations less than 10 kpc, but not in single-
nucleus post-merger systems, may indicate that the tails undergo a relatively 
unified burst of compact, clustered star formation during the 100 Myr period prior to 
the 
merging of the nuclei, and that no further local bursts occur. Conversely, many of the 
single nucleus systems (e.g. UGC 5101, IRAS 23365+3604) do have complex structure in 
their nuclear regions. Speculatively, 
this could be due to gas depletion in the tails; the denser nuclear environment 
may be able to fuel the bursts for much longer, or to fuel a sequence of many 
bursts as was indicated in Paper I. Fading of the starburst knots is very rapid at 
{\it U}\p; the knots fade from their peak intensity (which occurs at an age of 
30 Myrs for an instantaneous burst --- see below) by nearly 5 magnitudes by 
the time they are only 300 Myrs old. Fading by only 3 magnitudes would make 
most of the luminous knots observed undetectable. Since we detect no knots 
even more luminous than those seen here, it is therefore likely that we are 
detecting the knots near the peak of their luminosity and that most of the 
knots we observe at {\it U}\p are between 10-100 Myrs in age, which is borne 
out by an analysis of their ({\it U$-$B},{\it B$-$I}) colors (see below). Since 
the nuclear regions of the more dynamically advanced systems (i.e. Mrk 273, 
UGC 5101) continue to have luminous {\it U}\p knots, either these knots have an 
intrinsically higher peak luminosity (thus allowing them to be detectable for 
a longer period of time), the time-scale of the burst is much longer than the 
instantaneous burst modeled here, or there is an on-going generation of new 
knots. Note that these knots are shown below to be physically unrelated to the source of 
the majority of the bolometric luminosity and hence this result does not indicate 
anything about the evolution of the primary luminosity source in ULIGs. 

An alternative explanation to the dichotomy in the appearance of star-forming knots is 
that two entirely different populations of ULIGs exist. The above explanation implies 
that the ultraluminous systems remain ultraluminous throughout the merger process. 
Because we have selected our targets based on a luminosity criterion, it is possible that 
we have actually selected systems that are ultraluminous during the early merger 
phase and are accompanied by many star-forming extranuclear knots, and a different 
population that is ultraluminous during the late merger phase and which has no knots. 
One way to test this would be to search for relic knots in the tails of the single nucleus 
systems. If knots with colors similar to older stars were found in the single nucleus 
systems then this would demonstrate that during their earlier double-nucleus phase 
they appeared similar at U\p to the double-nucleus systems we are observing. However, 
we lack the sensitivity to detect these older knots.

Figures 3 \& 4 present surface brightness profiles for the warm and cool 
ULIGs, respectively, at U\p. The profiles were centered on the U\p flux 
centroid corresponding to the major ``nuclei'' (the brightest feature at B-band) 
identified previously in Paper I and Surace (1998). Circular azimuthally-
averaged isophotes were fitted using the IRAF/STSDAS task ELLIPSE. In each 
case the centers and ellipticities were held fixed. The radial profiles were 
characterized in two ways.

First, the radial profiles were fit with a \sersic (1968) surface brightness law:

\begin{equation}
I(r)=I_e exp (-b [(r/R_e)^{1/n}-1]) +I_{back}; \ \ \ b=1.9992 n - 0.3271
\end{equation}

\noindent which for n=4 is a de Vaucouleurs profile, and for n=1 is an exponential disk 
profile. With this definition of {\it b}, R$_e$ and I$_e$ are the half-light radius 
and intensity. The fitting was done with the IGOR data analysis package. In 
order to minimize the effects of seeing, which dominate the shape of the 
profile at small radii, the profiles were fit only at radii greater than the FWHM 
of the observed PSF (typically 1\arcsec). The \sersic profile has many degrees 
of freedom, and so additional constraints had to applied. I$_{back}$ was set 
equal to the observed background sky value far from the galaxy. I$_e$ was 
constrained to be greater than zero and less than the maximum observed 
surface brightness. Finally, R$_e$ was estimated by integrating the total light 
in successive radii until the half-light radius was found. Initially, the fitting 
algorithm allowed R$_e$ to vary as a free parameter; however, if it arrived at a 
clearly non-physical value (i.e. less than 100 pc or greater than 20 kpc), then 
R$_e$ was held fixed at the measured value. Even so, the derived value of {\it 
n} was very sensitive to the input parameters. This is primarily due to the 
distance of the ULIGs and the limited spatial resolution of the U\p 
observations. Zheng et al. \markcite{zheng} (1998) found, based on high spatial 
resolution {\it 
HST} data at I-band, that typical values of R$_e$ for a small sample of ULIGs 
were 1---6 kpc, with a median near 3 kpc. Unfortunately, this is similar to the 
radii over which the U\p radial profiles have been fit (typically 2---10 kpc). 
The \sersic profile has an inflection near R$_e$, and as a result {\it n} is poorly 
constrained (Figure 5). Furthermore, profiles with fixed parameters other 
than {\it n} are very similar at large radii for {\it n} $>2$. It is the profile at 
very small radii, which we cannot measure easily from the ground, that 
constrains the fit best. Only systems with apparently single nuclei were fit. 
This was because the uncertainties in estimating parameters in the presence 
of a second nucleus resulted in an unstable, and probably meaningless, value 
of {\it n}. The observed value of {\it n} is certain to within $\pm$0.3. Previous 
work has shown that {\it n} is usually slightly underestimated (Marleau \& 
Simard 1998).

The \sersic index {\it n} was found to lie between 0.8 (IRAS 01003-2238) and 3.5 (IRAS 
05189-2524), with a median value of 1.8. There is nothing to suggest a trend 
between the index {\it n} and far-infrared color. Both warm and cool galaxies 
have nearly the same range and median; Kolomogoroff-Smirnov statistics indicate that 
we cannot 
reject the null hypothesis that the two samples were drawn from the same 
parent sample (Press et al. 1992). The value of {\it n}$=$1.8 is intermediate between an 
exponential disk and a de Vaucouleurs profile with the same half-light radius, 
although it will more closely resemble the de Vaucouleurs profile than the 
exponential disk. The derived value of R$_e$ is contaminated by residual light 
from the QSO nucleus in I Zw 1, Mrk 1014, IRAS 07598+6508, and Mrk 231. In the 
remaining galaxies R$_e$ ranges from 1.9 kpc (IRAS 00091-0738) to 9.3 kpc 
(Mrk 273), with a median of 5.3 kpc. This is larger than that found by Zheng et 
al. \markcite{zheng} (1998). 

Additionally, for each galaxy, an r$^{-1/4}$ law profile was fit to the outer 
isophotes, beginning at a radius equal to the FWHM of the PSF. An exponential 
disk model was also fit. Similar to the result found above with the \sersic law, it 
is apparent that these two models fit the ULIGs to a varying degree (see Figures 3 \& 4). 
Those 
dominated by a bright central nucleus (I Zw 1, Mrk 1014, IRAS 07592+6508,  and 
Mrk 231) are all extremely well fitted by a de Vaucouleurs (\rqu) profile, while 
an exponential disk fits poorly. In some cases, both the exponential disk and 
the de Vaucouleurs profile fit the outer isophotes equally well (e.g. IRAS 
15206+3342, PKS 1345+12), while 
in other cases, the outer radii are well-fit by the \rqu profile and the inner regions by 
the exponential disk. Since the inner regions are more affected by dust extinction and 
resolution effects it is unlikely that this really represents the underlying stellar 
distribution.  There is no obvious trend between far-infrared 
colors (i.e., ``warmness'' or ``coolness'') and the type of best-fit model; it is not 
the case that all warm ULIGs have \rqu profiles, while the cool have 
exponential disks, which might have been expected if the two samples had different 
dynamical ages.

Since the morphologies of the systems are consistent with the remnants of 
major mergers, this is the expected result since relaxation of the disks results 
in their assuming a profile approximated by the r$^{-1/4}$ law. Significant deviations 
from an \rqu law are still seen--- 
specifically, they decrease more rapidly at large radii than might be expected for 
a de Vaucouleurs profile. Given the relatively sharp, discontinuous edges of 
the tidal features, which are the most extended parts of the merger systems, 
this turnover in the brightness profile at large radii probably represents it's 
truncation past the maximum extent of the tidal debris. This falloff in intensity 
is also similar to the behavior expected for an exponential disk, and may be due 
to incomplete relaxation of the merger remnant (Toomre \markcite{toomre} 1977). 

Similar results 
have been found for merging galaxies in general (Schweizer 1982, Wright et 
al. \markcite{wright}1990, Stanford \& Bushouse \markcite{stanford}1991), and more 
specifically for ULIGs at R \& K-
band (Wright et al. \markcite{wright}1991, Kim 1995). The latter is not surprising, 
since it is now 
known that most ULIGs are merging systems. What is perhaps surprising is 
that this result is true even at U\p and over a large range of radii extending to 
as far as 20 kpc. While longer wavelength observations (e.g. near-infrared) 
trace primarily the older stellar population and hence most of the stellar mass, 
the U\p observations should trace higher mass stars and sites of intense local 
star formation. However, an examination of Table 3 indicates that most of the 
compact features observed at U\p represent only a fraction of the total U\p 
flux and hence the U\p low surface brightness emission continues to trace the 
general galactic stellar population. This is similar to the result of Meurer et al. 
(1995). U\p traces stars with ages of at most a few hundred Myrs, which is similar to the 
dynamical merger timescale. The existence of an extended smooth stellar population at 
U\p may indicate that the ULIGs are currently undergoing spheroidal component 
growth, similar to that observed by Fanelli et al. (1997) \markcite{fanelli} and Smith et 
al. (1996) \markcite{smith} in the far-ultraviolet in two nearby starburst galaxies.

\subsection{Colors and Derived Starburst Ages}

\subsubsection{Modeled Colors}

Table 3 presents the photometric data for the U\p observations, all of which have been 
transformed to the Johnson U-band from the observed U\p 
filter.
The colors of the observed features were compared to various modeled colors. The 
U\p colors provide essentially no discriminant between AGN and starburst 
activity. The synthetic optical QSO model introduced in Paper II predicts ({\it 
U$-$B, B$-$I}) colors for a QSO at the sample median redshift of ($-$0.78, 0.67) and the 
UVSX sample of Elvis et al. (1994) has colors of ({\it U$-$B}),({\it 
B$-$I}) = ($-$0.82,1.12)$\pm$(0.16,0.19), 
which are similar to a modeled instantaneous starburst with an age of 10 Myrs. As 
discussed in Paper II, long 
wavelength emission ($>$ 2 $\mu$m) therefore remains the best possible broad-band 
color 
discriminant between AGN and recent star formation activity. Throughout this 
analysis we will continue to consider the U\p emission as a result of either 
starburst activity or AGN depending on the results found at other wavelengths 
(Surace 1998). Larson \& Tinsley \markcite{larson} (1978) find typical (U$-$B) colors 
for non-
interacting galaxies in the RC2 in the range of $-$0.2 to  $+$0.6, and Guiderdoni et 
al. (1988) find typical values of ({\it U$-$B}) for spirals of 0.03. Such a wide 
span of colors is probably due to the variable star formation histories of the 
``normal'' galaxies involved. Perhaps a more meaningful comparison can be 
made to the ({\it U$-$B}) colors of individual stellar spectral types: ({\it U$-
$B}) = $-$1.15, $-$1.06, $-$0.02, 0.07, 0.05, 0.47, and 1.28 for spectral classes O5, 
B0, A0, F0, G0, K0, and M0, respectively (for stars on the main sequence --- the 
corresponding giant stars are somewhat redder; Allen \markcite{allen}1973). Only OB 
stars can 
directly contribute to ({\it U$-$B}) colors significantly less than 0.

The derived colors are compared to evolutionary synthesis models. Figure 4 
illustrates the (U$-$B, B$-$I) colors derived from two of the most popular 
models: BC95 (first described in Bruzual \& Charlot 1993; the updated version is 
BC95) and Starburst 99 (Leitherer et al. 1999). The BC95 models represent 
isolated, bare stellar ensembles (photospheres), whereas the Starburst 99 models also 
include 
nebular continuum emission, which has an effect at both very short ($<$ 4000\AA) and 
very long ($>$ 1.5 $\mu$m) 
wavelengths (Leitherer \& Heckman 1995). BC95 extends to much greater ages than 
Starburst 99, but it is 
unlikely that any of the star-forming regions observed here are so old. 
Neither model includes the effects of dust, which will both redden the 
intrinsic colors and also deplete the nebular continuum. In this paper most of 
the ages will be based on the BC95 models, since it is easier to understand how 
processes such as reddening affect the observed colors when applied directly 
to stars without additional complications such as nebular continuum which 
may not be well-understood. 

The age estimates based on color diagrams of this sort are 
relatively insensitive to the exact shape of the IMF, which in this case is a Salpeter 
initial mass  function (IMF) with upper and lower mass cutoffs of 125 and 0.1 
M$_{\sun}$, respectively.. This is because the colors 
of a stellar (starburst) ensemble are dominated by the most luminous stars 
since what is actually observed is a luminosity-weighted average of all the 
stars in the burst (Leitherer 1996). It would require very extreme changes in 
the mass index of the IMF in order for late-type stars to dominate the colors by sheer 
numbers. The low mass end of the IMF primarily acts as a reservoir for the 
mass of the burst; most of the luminosity is emitted by only a small ($\approx$ 
10\%) mass fraction of the burst. Truncation of the lower end of the IMF, 
therefore, primarily adjusts the luminosity per unit mass of the burst, and 
thus the rate at which the gas supply is consumed. For example, truncating the 
IMF at 3 \msun in the BC95 model above only increases the bolometric 
luminosity per unit mass by a factor of 5 and leaves the color evolution almost 
unchanged until roughly 1 Gyr (at which time the entire stellar population 
has left the main sequence and therefore becomes very red; Charlot et al. 
(1993)). Similarly, adjusting the upper mass cutoff from 120 to 60 \msun leaves 
the color evolution unchanged after the first few million years due to the very 
rapid evolution of the upper end of the IMF. This model is thus likely to be a 
good one for a typical long-lived stellar population extending from the most 
massive stars to the least massive. Changing the upper and lower mass cutoffs 
is likely to only change the bolometric luminosity per unit mass, and even 
then only by a factor of a few. 

The short wavelength colors are most strongly 
affected by dust and metallicity.  Figure 7 presents different metallicity models from 
Starburst 99 (BC95 assumes solar metallicity). Shown are solar metallicity, twice solar 
metallicity, and a 0.05 solar metallicity case. Note that the solar and twice-solar
metallicity models have nearly the same colors. Given that the local ULIGs 
appear to be collisions of mature, gas-rich spiral galaxies (Sanders et al. 
1988a), it is not implausible that the gas is of near solar-metallicity. Unlike the situation 
in the near-
infrared (i.e., as in Paper II) direct thermal dust emission cannot affect the short 
wavelength colors as these wavelengths correspond to temperatures far above 
the dust sublimation temperature. This leaves 
reddening as the dominant source of color variance. 

Two dust reddening models are considered. The first is a simple foreground 
screen model, where all the dust lies on a path between the emitting source 
and the observer. The reddening law of Rieke \& Lebofsky \markcite{rieke} (1985) was 
used. 
This is illustrated in Figure 8 with a solid line vector corresponding to an 
extinction of 1 magnitude at V-band. Ages for the starbursts are derived by 
dereddening their observed colors until they intersect the modeled colors for a 
solar metallicity starburst. Fortunately, objects can only be made more red 
by extinction, and thus the colors can define upper limits to the stellar ages. However, 
while this may be a usable extinction 
model in the near-infrared, at wavelengths as short as U\p scattering may 
play an increasingly important role. Witt et al. (1992) have shown that for 
some scattering geometries it is difficult to achieve effective extinctions at U-
band much greater than a few magnitudes even when the line of sight 
extinction is as high as 25 magnitudes. The ``dusty galaxy'' model of Witt 
et al. (1992) in which dust and stars uniformly fill a spherical volume was used, and 
which was
also used in the multi-color analysis of Surace et al. (1999).
This is motivated by the observed morphology of star-formation in the nuclear regions 
of these galaxies (Surace et al. 1998,1999). Hill et al. (1995) find that the Witt et al. (1992) 
``dusty nucleus'' model provides a better fit to ultraviolet observations of HII regions in 
M81. However, this geometry is more appropriate to individual HII regions and not to 
the galaxy as a whole. In any case, for the maximum possible effective extinctions 
allowed by the U\p observations the dusty nucleus and dusty galaxy models have nearly 
identical broadband color properties. 
Figure 8 illustrates 
the reddening path of the ``dusty galaxy'' model in units of total optical depth at V-band 
of 
$\tau$=0, 5, 10, and 15. Qualitatively it is very similar to the foreground 
extinction model. However, the reddening vector for the dusty galaxy model is 
slightly flatter due to the decreased effective optical depth at U\p as a result of 
the increased scattered light component relative to B and I-band. As a result, 
derived ages based on this model will typically be 20\% higher than those 
based on a foreground screen. 
All ages quoted here are upper limits based on the foreground dust extinction 
model, since it is simpler in application.

\subsubsection{Observed Colors}

The integrated ({\it U$-$B}) colors of the ULIGs span a very large range from -0.5 
(IRAS{\ts}12112+0305) to 1.1 (UGC{\ts}5101), with a median of 0.4. Such an 
enormous range in integrated colors is similar to the Larson \& Tinsley 
\markcite{larson}(1978) result, and probably does not constrain the galaxies' stellar 
populations 
significantly. Due to the wide range in colors, another approach 
must be used. It is more meaningful to address the colors of the individual 
compact high surface brightness features observed in the ULIGs, since these 
features correspond to the young starburst component of the galactic stellar 
population. Figure 8 shows the colors of the compact, high surface brightness 
nuclear regions of the ULIGs presented in Table 3.

The lack of resolved U\p knots among many of the cool ULIGs forces an 
examination of their ``nuclear'' colors. In Surace (1998) it was shown that the 
``nuclear'' regions (defined by a diameter of 2.5 kpc) typically had optical/near-
infrared colors similar to young star-forming regions with contributions from 
hot dust and mild ({\it A}$_{\rm V}$= 2--4 magnitudes) of reddening. The median 
nuclear 
colors for the cool ULIGs observed at U\p are ({\it U$-$B, B$-$I})=(0.1, 2.1), 
with ({\it U$-$B}) ranging from 1.2 to $-$0.6. The median age corresponding to 
the colors of the individual nuclei are 44 Myrs and 35 Myrs for the BC95 and 
LH99 models, respectively. Although it is not possible to discriminate between 
AGN and starburst activity based on UBI colors alone, the lack of a K\p excess 
in the cool ULIGs (Surace 1998) supports the interpretation of the blue (U$-$B) 
colors as being indicative of young stars. The presence of young stars in the 
cool ULIG nuclei (and in particular the single nucleus systems such as UGC 
5101) indicates the presence of on-going young star formation, in contrast to 
the tidal tails ---if a burst has occurred there it has already faded to the point 
of undetectability. In several cases U\p knots are resolvable within a radius of 
4$-$5 kpc of the nucleus. In IRAS 12112+0305 the knots flanking the northern 
nucleus as well as the knots in the southern arc have dereddened ages of 34 
and 6 Myrs, respectively. The southern knot in Mrk 273 has colors most similar 
to a 1 Gyr old burst (although this seems unlikely due to the fading discussed 
above), while the western knot appears to be about 200 Myrs. Finally, IRAS 
23365+3604 has knots ranging between 3--60 Myrs.

Examining the ``warm'' ULIGs is more problematic, since there is evidence for 
AGN activity in the nuclei of all of these systems (Sanders et al. 1988b, Papers I 
\& II), and since AGN have ({\it U$-$B,B$-$I}) colors very similar to those of a 
young starburst. It is therefore likely that any examination of the nuclear 
(2.5 kpc diameter) regions will simply reflect the presence of AGN light. 
Instead, we break the warm ULIGs into several subcategories. 

The double nucleus systems have the most complex morphology.
Of the 3 systems 
with double nuclei, IRAS 08572+3915 and Mrk 463 each have many U\p knots, 
while Pks 1345+12 seems to have few. Although this latter case may be a result 
of the low S/N of our observations, the {\it HST}  data (Paper I) also shows little 
in the way of knot structure as well. The knots in the tails of IRAS 08572+3915 
appear to have dereddened ages of 5--10 Myrs, while those in Mrk 463 vary 
from 4-90 Myrs, although the median is only 6 Myrs. The stellar galactic nuclei 
(based on imaging at other wavelengths) IRAS 08572+3915e and Mrk 463w 
have ({\it U$-$B, B$-$I}) colors of ($-$0.3, 1.6) and ($-$0.5,1.5) yielding 
dereddened ages of 45 and 25 Myrs, respectively. Similarly, it was shown in 
Paper II that the optical emission from IRAS 08572+3915w is predominantly 
starlight, since the AGN only begins to contribute at near-infrared 
wavelengths. For it we derive a starburst age of 37 Myrs. Pks 1345+12 has 
integrated colors so red ({\it U$-$B, B$-$I})=(0.8, 2.5) as to be similar to a very 
old, late-type stellar population, although this could also be the result of 
{\it A}$_{\rm V}$ =  2--3 magnitudes of intervening dust. 

The single nucleus systems have simpler morphology with fainter tidal features.
The integrated colors of 
IRAS{\ts}01003$-$2238 indicate a maximum age of 40 Myrs, although at least 
some of the knots observed with WFPC2 are at most 5 Myrs in age (Paper I). Of 
the single-nucleus non-QSO systems, Mrk 231, IRAS 12071$-$0444, and IRAS 
05189$-$2524 have the most well-developed tidal structure, but only the former 
two have any evidence for resolvable starburst knots. This is probably due to 
the lower spatial resolution at U\p; there is evidence in the WFPC2 data (Paper 
I) for an extended starburst region in the nucleus of IRAS 05189$-$2524, but 
which cannot be resolved here. The estimated age of the southern, presumably 
stellar, region of IRAS 12071$-$0444 is 125 Myrs. The northern region in the 
vicinity of the putative active nucleus contains several knots of star formation 
which we cannot resolve here, and which may be quite young. IRAS 15206$-
$3342's western half has colors of ($-$0.4, 1.5), which are similar to those of a 
20 Myr old burst. The eastern component shows a peculiarly large U\p excess 
($-$0.9, 1.6), which may be indicative of contamination by the putative active 
nucleus (Paper I). The same is likely to be true of the nuclear colors of IRAS 
05189$-$2524, whose dereddened colors are similar to a 20 Myr old starburst. 
Mrk 231's ``horseshoe'' has dereddened colors similar to a starburst 225 Myrs 
old. The extended structure in the QSO I Zw 1 appear similar to the spiral arms 
of a spiral host galaxy. Mrk 1014 has several knots embedded in its tidal tails, 2 
of which were detected with WFPC2. Unfortunately, they are sufficiently faint 
that their colors are not well determined; they are likely to be in the age range 
3--30 Myrs. Lastly, the extended features to the south and east of IRAS 
07598+6502 were not obviously detected; however, these features were of fairly 
low effective surface brightness due to the seeing. A careful examination of 
the region as a whole indicates ({\it U$-$B}) similar to a stellar population of 
5--10 Myrs. 

It is apparent from Figure 8 that most of the ``warm'' ULIGs have bluer nuclear colors 
than the ``cool'' ULIGs. However, their dereddened ages are extremely similar. 
Thus, while the redder colors might be indicative of greater age, it is more 
likely that they indicate a greater degree of dust obscuration, either by virtue 
of more line-of-sight dust or by having a dust geometry less favorable to 
scattering and thus a greater effective optical depth. This would be consistent with the 
evolutionary scenario proposed 
by Sanders et al. (1988a), in which dust is cleared away from the AGN in the 
warm ULIGs; presumably the same process would clear dust away from the star 
formation regions as well. The broad-band colors do not show a 
detectable maximum age difference between the two samples of ULIGs (although this is 
poorly 
constrained ). This requires that the dust clearing time be relatively short 
compared to the timescale of the nuclear burst ($\approx$ 100 Myrs) in order for the 
dust geometry to be the dominant color determinant above and beyond the effects of 
stellar aging. 

Figure 9 illustrates the nuclear colors of single vs. double nucleus ULIGs. 
Unfortunately, several systems known to have double nuclei at other wavelengths 
either could not be resolved or had confused detections at one or more wavelengths 
(Pks 1345+12, IRAS 22491-1808, IRAS 12112+0305, IRAS 0119-2307). As a result the 
density of points in the figure is deceptive. Although the single nucleus systems are of 
necessarily greater dynamical age, this is not clearly reflected in their nuclear colors 
which have similar dereddened ages for both types of systems. This would argue against 
the evolutionary scenario, since it implies that dynamical age is not linked to dust 
reddening, which appears from the above to be linked to the mid-infrared colors of the 
ULIGs. However, the blue double nucleus point are strongly affected by IRAS 
08572+3915 and Mrk 463. Since exclusion of these would produce different results, it is 
likely that there are too few galaxies in the sample to as yet draw meaningful 
conclusions.

\subsection{Luminosities}

\subsubsection{Observed Total U-band Luminosities}

The total luminosities $\nu$L$_{\nu}$ at U-band were computed using the 
conversion that a magnitude zero star has a flux density F$_{\nu}$ of 1810 Jy 
in the U filter (Bessell 1979). The average rest-frame luminosity (i.e. the total 
amount of energy in units of 3.9$\times$10$^{33}$ ergs s$^{-1}$ in the U-band 
filter) is \nul = 10$^{10.7}$ \lsunns , while the median is 10$^{10.0}$ \lsunns. 
This high average value is strongly influenced by the nuclear emission of the 
QSOs I{\ts}Zw{\ts}1, Mrk 1014, and IRAS 07598+6508. Excluding these three 
systems yields an average that is also \nul = 10$^{10.0}$ \lsunns . In terms of 
absolute magnitude, the median is {\it M}$_{\rm U}$= $-$20.8 and the average 
is $-$21.3. These results are consistent with the values extrapolated from the 
far-UV observations of IRAS 22491$-$1808 and IRAS 12112+0305 (Trentham et 
al. \markcite{neil} 1999). Given that infrared-selected QSO host galaxies are typically 
similar in 
luminosity to their nuclei (McLeod et al. 1994, Surace 1998), the ULIG QSOs are 
likely to have hosts similar in luminosity to the mean luminosity of the ULIGs.

\subsubsection{Contribution of U-band structure to the bolometric luminosity}

We consider the probable bolometric luminosity of the observed knots and nuclear 
regions. The {\it U}-band bolometric correction (BC) is relatively robust against 
changes in the IMF mass cutoffs, since the energetics are dominated by the most 
massive stars. Indeed, adjustments to the IMF mass cutoffs, while changing the 
bolometric luminosity per unit mass by a factor of a few, have little effect on color as 
explained above, and hence the bolometric 
correction remains unchanged. Throughout the starburst age range of 0-1 
Gyr, the bolometric correction at U is predicted by both BC95 and Starburst99 to 
vary only from about 0.5 to 2, and is very nearly 1 during the span of 10-600 
Myrs for the modeled starburst. The likely bolometric luminosity of the star 
formation observed at U\p was determined by applying the bolometric 
correction applicable to a given measured emission region based on it's 
estimated dereddened upper age limit, which was in turn based on its {\it UBI} 
colors. This was then increased by dereddening by the amount indicated by the 
{\it UBI} colors (typically A$_V$ = 1--1.5 magnitudes). These 
estimates were all based on a foreground dust extinction model.

The cool ULIG bolometric luminosities due to current star-formation were calculated by 
considering the 
emission inside the nuclear (2.5 kpc diameter) regions which are known optically to 
host complex star-forming knots which we cannot resolve here, as well as any 
observed 
knots outside this region. The estimated nuclear {\it L}$_{\rm bol}$ ranges from 
10$^{9.7}$\lsun 
(UGC 5101) to 10$^{11.3}$\lsun (IRAS 00091$-$0738), with a median of 
10$^{10.7}$\lsun and 75\% lying in the range 10$^{10.4}$--10$^{10.9}$\lsun. 
The star-formation observed at {\it U}\p thus accounts for approximately 3\% 
of the bolometric luminosity in a typical cool ULIG, assuming an average {\it 
L}$_{\rm bol}$ of 10$^{12.3}$\lsunns (Sanders et al. 1988a). This is extremely 
similar to the result derived by Surace (1998) from optical/near-infrared 
wavelengths.

The warm ULIGs' star-formation budget was tallied by adding up all the 
emission in the observed knots as well as non-AGN nuclear regions (i.e., IRAS 
08572+3915e) and all the nuclear regions where star-formation is believed to 
dominate at U\p based on WFPC2 B-band imaging (e.g., IRAS 08572+3915w, IRAS 
12071$-$0444). The warm ULIG bolometric luminosities vary from 10$^{9.9}$ 
(IRAS 08572+3915) to 10$^{11.7}$\lsunns (IRAS 15206+3342), with a median of 
10$^{10.6}$\lsunns . Again, this is similar to results found at other 
wavelengths. The warm ULIGs (not including 3c273) have a mean {\it 
L}$_{\rm bol}$=10$^{12.26}$\lsunns; the average warm ULIG has a 
contribution of $\approx$2\% by star formation observed at U\p to the 
bolometric luminosity, although this number is as high as 30\% in at least one 
case (IRAS 15206+3342).

Estimation of the bolometric luminosity is more problematic in the presence of 
mixed stars and dust with scattering. Witt et al. (1992) point out that while the 
effective optical depth at short wavelengths like U\p may be quite low due to 
scattering, the true optical depth may be many times higher. As a result, while 
the galaxy may be quite luminous at U\p, a significant amount of longer 
wavelength emission is absorbed and reradiated in the thermal infrared due to 
the high total optical depth. In this case, the colors of the ULIGs are consistent 
with young stars mixed with dust mixed and having a total V-band optical 
depth of 5---20. The ``dereddening'' correction will increase the bolometric 
luminosity by a factor 4---10$\times$ greater than that of the foreground 
screen alone (Witt et al. 1992), and the observed star formation could 
contribute a sizable fraction of the known bolometric luminosity of the 
systems. This is unlikely, however. If the observed extended star formation did 
contribute a significant fraction of the bolometric luminosity, then the 
galaxies would be extended at every wavelength in a manner similar to that at 
U\p. However, it is known that the ULIGs become increasingly more compact 
in the near-infrared (Surace et al. 1999, Scoville et al. \markcite{scoville2} 1999) and 
ultimately are 
unresolved from the ground at mid-infrared wavelengths (Soifer et al. 1999). 
Since the latter are near the peak of the spectral energy distribution, this implies that 
the star-forming 
knots seen at U\p are not significant contributors to the bolometric 
luminosity, and that the dusty galaxy model is not a good approximation to the 
actual dust distribution.

\section{Implications for High-Redshift Systems}

Our 3410\AA \ observations of low-redshift ULIGs are applicable to 
observations of high-z galaxies at optical and near-infrared wavelengths. In 
particular, rest-frame U\p observations correspond to R-band observations at 
{\it z}=0.88 and H-band at {\it z}=3.9. Recent observations of the Hubble Deep 
Field (HDF) with ISO (Rowan-Robinson et al. 1998) have shown that the 
preponderance of strong mid-IR emitters are apparently interacting, star-
forming galaxies. Deep imaging surveys with the Sub-mm Common User 
Bolometer Array (SCUBA) instrument on the James Clerk Maxwell telescope 
(JCMT) on Mauna Kea have recently discovered large numbers of sub-mm 
luminous high redshift galaxies (Smail et al. 1997). The projected star-
formation rates for these galaxies are as high as 100 \msunns/year; the ULIGs 
are the only local analog of starbursts with such high star formation rates 
(Hughes et al. 1998, Barger et al. \markcite{amy} 1998). Like the high redshift sub-mm 
galaxies, 
ULIGs are also found in interacting systems, and have SEDS with extremely 
strong far-IR components. Therefore, a more thorough understanding of the 
properties of ULIGs is likely to be the best opportunity for understanding the 
properties of these systems at high redshift. 

The local ULIGs are observed to have angular diameters as large as 30 kpc. 
If the high redshift sub-mm galaxies are mergers of L$^*$ galaxies similar to 
the low redshift ULIGs, then they will still be resolvable from the ground with 
diameters of 5\arcsec \ at {\it z}=1 and 6\arcsec \ at {\it z}=3 for a q$_0$=0.5 
universe (Peebles \markcite{peebles} 1993). For q$_0$=0 these are 4\arcsec \ and 
3.5\arcsec. This is 
similar in many cases to the observed angular sizes of the optical counterparts 
of the sub-mm galaxies (Smail et al. 1998). The observed surface brightness of 
the galaxies is independent of q$_0$ and varies as (1+{\it z})$^4$. The apparent surface 
brightness of an ULIG at {\it z}=0.1 decreases by 2.6 magnitudes at {\it z}=1 and 
5.6 magnitudes at {\it z}=3. The extended tidal structures in the ULIGs have 
typical surface brightnesses of 21-23 magnitudes per square arcsecond at U\p. 
The equivalent depth at {\it z}=1 in R-band ($\approx$ 25---27) is easily 
reached in less than an hour by Keck (Hogg et al. \markcite{hogg} 1997). The brightest 
U\p 
extended tidal features at {\it z}=3 will have surface brightnesses approaching 
H=27, and are probably unattainable by ground-based telescopes. However, the 
new generation of 8m class telescopes, equipped with adaptive optics, will have 
angular resolutions of approximately 40 milliarcseconds at {\it H}. This will 
provide a physical spatial resolution at high redshift comparable to or higher 
than that achievable for local ULIGs now from the ground at U\p. The star-
forming clusters, which have typical HST resolved sizes of 100 pc (Surace et al. 1998), 
will still be 
unresolved. The brightest of these unresolved knots, as well as the nuclei, will 
approach {\it H}=25---26, and will be detectable by a large telescope in 1---2 
hours (due to the enormous increase in point source detectability afforded by 
high spatial resolution imaging in background-limited situations). If the star-
forming galaxies now being seen at high redshift are a result of collisions and 
mergers, then future optical and near-infrared observations of sub-mm 
luminous star-forming galaxies at high redshift are likely to appear similar in 
appearance to the U\p observations of ULIGs, with tidal debris and clustered 
star formation clearly visible. The Smail et al. (1998) results do actually 
indicate that 6/16 optically identified high-z galaxies are interacting systems 
similar in appearance to local ULIGs. The ULIG mergers appear in many cases 
to also fuel an AGN; the rate of apparent AGN activity in high redshift submm-
luminous galaxies might indicate whether the merger process, while efficient 
in fueling a black hole, is also capable of creating new ones. Current estimates 
of the rate of optically detected AGN in high redshift sub-mm galaxies are 
$\approx$ 20\%. This is similar to the fraction of ``warm'' AGN-like systems 
seen in complete samples of ULIGs in the local universe (Kim 1995).

Observers will attempt to morphologically classify their observations of high 
redshift galaxies.
Initial observations at high redshift may be of relatively poor spatial 
resolution, making it difficult to discern the high surface brightness features 
seen in many of the images of local ULIGs. Examinations of the radial profiles 
of objects at high redshift can help discern if they are ``elliptical-like'' or 
``disk-like''. As shown in \S 3.1, if the high redshift systems are similar to 
those examined here, then they will predominantly have radial profiles 
intermediate between the two. However, it may be possible to 
image high redshift star-forming galaxies and determine if their detailed 
morphologies are consistent with the local merger ULIGs observed here. Preliminary 
analyses of high-redshift galaxies believed to be undergoing active star-formation (for 
example, in the Hubble Deep Field) do show evidence for \rqu profiles, possibly due to 
the same merger activity seen in the local ULIGs.

\section{Conclusions}

\noindent 1. Many ULIGs are luminous at U\p (median \nul = 10$^{10.0}$ \lsunns) and 
have evidence for 
considerable small-scale structure. This structure is almost identical to that 
seen at optical wavelengths.

\noindent 2. Many of the double-nucleus systems have luminous knots of U\p 
emission in their tidal tails, which are absent in the single-nucleus systems. 
This may indicate that the clustered star-formation in the tails is predominantly 
brief and occurs early in the merger process prior to the coalescence of the nuclei.

\noindent 3. The star-forming knots in the cool and warm ULIGs have similar upper 
age limits (10--100 Myrs), although those in the warm ULIGs are generally consistent 
with less dust obscuration. This is consistent with greater dust clearing if the warm 
ULIGs are older systems than the cool.

\noindent 4. The bolometric luminosity of the star-formation observed at U\p 
is generally only a very small fraction ($\approx$ 3\%) of the known 
bolometric luminosity, a result nearly identical to that derived at optical and 
near-infrared wavelengths. This implies that even over a span of a factor of 20 
in visual extinction, the star-forming structure remains similar. Most of the 
clustered star-formation we can detect at optical and even at near-infrared 
wavelengths 
probably occurs in knots that are either not heavily extinguished along the line of 
sight or have a favorable scattering geometry. The source of most of the bolometric 
luminosity remains completely hidden at these wavelengths, a result directly supported 
by mid-infrared imaging (Soifer et al. 2000).

\noindent 5. The radial profiles of the ULIGs at U\p appear to be similar to a \sersic 
profile with index {\it n} = 1.8. This is an 
intermediate form between the exponential disk and the \rqu law. This is the 
expected result for a partially relaxed merger remnant.

\noindent 6. If they are similar to the local ULIGs, high-redshift ULIGs 
observed at optical wavelengths will have resolvable tidal structure and star-
forming features which can be detected with ground-based telescopes.

\acknowledgements

We thank Gerry Luppino and John Tonry for the construction of the Orbit 
camera, without whose remarkably high quantum efficiency these 
observations could not have been made. We also thank John Dvorak and Chris 
Stewart for their skill at operating the telescope. We thank Lee Armus and B.T. 
Soifer for providing valuable discussions on the ULIGs, Aaron Evans for his 
useful comments on early drafts of this text, and Alan Stockton for advocating 
the value of U-band observations. 
We thank an anonymous referee whose comments helped strengthen this presentation.
J.A.S. was supported by NASA grant NAG 5-3370 and 
by the Jet Propulsion Laboratory, California Institute of Technology, under contract 
with NASA. DBS was supported in part by JPL contract 
no. 961566.

\clearpage
\appendix
\section{Notes on Individual Objects}

The following notes are provided as a description of the individual targets and 
to aid in understanding the relationship between the {\it U}\p emission and 
features seen at other wavelengths.

{\it IRAS 00091$-$0738} --- the {\it U}\p emission originates primarily in the 
stellar ``nucleus'', with very little emission coming from the dense tidal tail to 
the south.

{\it I Zw 1} --- the {\it U}\p emission originates almost entirely (90\%) in the 
Seyfert 1 nucleus. Both tidal arms can be traced at {\it U}\p and closely mirror 
the {\it B}-band morphology. In particular, the two condensations in the NW 
arm and the bright blue condensation in the SW arm are both easily detected at 
{\it U}\p. 

{\it IRAS{\ts}01003$-$2238} --- no features are discernible, which is expected 
from the small size scale of the extended structure seen in the WFPC2 images 
(Surace et al. 1998) and the poor ground-based resolution. It's colors are 
similar to a starburst of age 10 Myrs.

{\it IRAS 01199$-$2307} --- both nuclei are detected at {\it U}\p, but the 
emission is dominated by the SW nucleus by roughly 3.5:1. The tidal arms are 
undetected at {\it U}\p.

{\it Mrk 1014 = PG 0157+001} --- the eastern tidal arm is very prominent at {\it 
U}\p, breaking into many smaller clumps and knots. The QSO nucleus accounts for 
about 
86\% of the total {\it U}\p emission. The integrated host galaxy color is {\it U$-
$B}=0.72, similar to that of old stars.

{\it IRAS{\ts}05189$-$2524} --- the {\it U}\p emission essentially mirrors that 
at all
other wavelengths (Surace et al. 1998, 1999). The EW and NS tidal loops are easily seen. 
There are no apparent knots, although there may be some undetected in the resolved, 
extended nucleus.

{\it IRAS{\ts}07598+6502} --- the star-forming knots detected at {\it B} are 
undetected at {\it U}\p. The nucleus accounts for 94\% of the emission.

{\it IRAS{\ts}08572+3915} --- large knots of emission trace the leading edges of
all the tidal structures for their entire length. Many small knots are seen
in the vicinity of the putative NW active nucleus. The SW nucleus dominates
 at short wavelengths, perhaps indicating that a strong starburst
is taking place there. Both nuclei have ({\it U$-$B}) $< -$0.25 in a 2.5 kpc 
diameter aperture; this, combined with their ({\it B$-$I}) colors from Surace et 
al. (1998), indicates the presence of stars that cannot be much more than 100 
Myrs old in both nuclei.

{\it UGC{\ts}5101} --- the {\it U}\p emission comes from a complex extended region in 
the core
of the merger system. The tidal tails are barely detectable.

{\it IRAS{\ts}12071$-$0444} --- clearly extended in the same manner as the 
WFPC2 images. Nearly all the compact {\it U}\p emission arises from the 
northern nucleus seen at {\it B}. The southern complex of knots is clearly 
detected at {\it U}\p.

{\it IRAS{\ts}12112+0305} --- all of the tidal structure seen at {\it B} \& {\it I} is 
seen at {\it U}\p. The condensations in the center of the southern tidal arc are 
clearly visible. The galactic ``nucleus'' in the northern part of the system is 
dominated by a compact knot of emission at {\it U}\p, with another knot 
located on the short tidal arc extending to the west. Most notable by it's 
absence is the star-like nucleus in the center of the system, which has 
completely disappeared between {\it U}\p and {\it B}-band, implying a line-of-
sight extinction greater than 3 {\it A}$_{\rm V}$. This may explain the 
appearance of the galaxy shortward of 2200 \AA, since the Trentham et al. 
\markcite{neil} (1999) HST/FOC data indicate almost no detectable  emission from the 
two 
nuclei themselves.

{\it Mrk{\ts}231} --- The {\it U}\p emission comes almost entirely from the 
active nucleus and from the southern star-forming ``horseshoe''. The 
horseshoe has ({\it U$-$B})=0, yielding a maximum age of a few hundred Myrs. 
The corkscrewing structure extending from the NW of the nucleus and 
wrapping around through the west to the extreme south of the nucleus is also 
detected at {\it U}\p.

{\it Mrk{\ts}273} --- Although the tidal tails are clearly detected at {\it U}\p, 
nearly all the emission arises in the central few kpc, which contains several 
large knots of {\it U}\p emission which are not seen in the WFPC2 I-band 
images (Surace 1998) or in CFHT K-band images (Knapen et al. \markcite{knapen} 1997) 
In 
particular, the extremely prominent southern U\p knot does not appear in 
either of these images. The ``nucleus'' is the luminous source to the north.

{\it Pks{\ts}1345+12} --- this galaxy is very red ( {\it U$-$B}=0.73, {\it B$-
$I}=2.54) and hence difficult to detect with high S/N at {\it U}\p. However, it 
appears that most of the {\it U}\p emission arises in the vicinity of the 
western nucleus.

{\it Mrk{\ts}463} --- the eastern nucleus is distinctly elongated N-S, and the 
western nucleus is elongated E-W exactly as expected from the WFPC2 images of 
Paper I. Nearly all of the U\p emission in the eastern nucleus comes from the 
position of the northern ``knot'', and not the southern infrared ``nucleus'' 
(Surace \& Sanders 1999), which is consistent with its interpretation as an AGN 
ionization feature. Although the eastern nucleus is physically smaller, and 
hence has a higher peak surface brightness, it is the western nucleus that 
dominates in overall luminosity at U\p. All of the ``knots'' identified in the tidal 
ring structure are extremely luminous at U\p. Their colors indicate ages less 
than 10 Myrs.

{\it IRAS{\ts}14348$-$1447} --- again, the U\p emission mirrors that at B. The 
blue knots seen at B surrounding the southern nucleus contribute heavily to 
the U\p emission in that nucleus, while the northern nucleus has a chain of 
knots in the base of the northern tail that contribute to the U\p emission.

{\it IRAS{\ts}15206+3342} --- closely resembles the WFPC2 optical images 
(Paper I). Most of the U\p emission seems to originate in the compact nuclei in 
the southern half of the primary ``string'' of knots, consistent with their blue 
colors.

{\it IRAS{\ts}15250+3609} --- the high surface brightness central galaxy 
component is well detected at U\p, as are the tidal features. Surprisingly, what 
appeared to be two separate features at I-band now appear to be a single arc 
bending away from the galaxy to the NE. The nearly circular arc to the south is 
undetected. The U\p emission is emitted by a compact source in the southern 
half of the central galaxy nucleus.

{\it IRAS 22491$-$1808} --- most of the U\p emission originates in the western 
nucleus and northwestern tidal arm, and corresponds to the star clusters seen at 
optical wavelengths. Trentham et al. \markcite{neil} (1999) find that most of the 
emission 
shortward of 2200 \AA originates in the arm.

{\it IRAS 23365+3604} --- the U\p emission closely mirrors that at B. All of the 
knots detected at B are luminous at U\p, and have derived (U$-$B) $<$0, 
indicating probable ages of 10 Myrs or less. The tails are almost undetectable; 
the most prominent extended emission at U\p is the arclike extension of the 
disk to the NW of the nucleus. Overall, the total integrated galaxy colors are 
similar to those of an old stellar population.

\clearpage

\figcaption{U\p images of ``warm'' Ultraluminous Infrared Galaxies. The 
images have been smoothed by convolution with a 0.2\arcsec gaussian kernel. Minor 
ticks are 1\arcsec, major ticks are 5 \arcsec, and the scale bars represent a 
physical scale of 10 kpc. Logarithmic contours have been added to illustrate 
structure in the saturated regions of the grayscale figure.}

\figcaption{Same as Figure 1, but of ``cool'' Ultraluminous Infrared Galaxies. }

\figcaption{Azimuthally averaged radial profiles of ``warm'' ULIGs at U\p. 
Vertical axis is surface brightness in magnitudes per square arcsecond, while 
the horizontal axis is radius$^{1/4}$ in kpc. In this form, an r$^{-1/4}$ law is a 
straight line. Solid lines are the best fit de Vaucouleurs profile to the outer 
regions of the profiles, while the dashed lines are best fit exponential disks.}

\figcaption{Same as Figure 3, but presents averaged radial profiles of ``cool'' 
ULIGs at U\p. }

\figcaption{\sersic surface brightness profiles for a range of index {\it n} with a half-
light radius of 2 kpc, typical of many of the systems examined here. Profiles 
with {\it n}$>$2 are very similar at large radii.}

\figcaption{Comparison between Bruzual \& Charlot BC95 (dashed line) and Leitherer 
et al. Starburst 99 (solid line) models. Shown are the (U$-$B,B$-$I) colors for an 
instantaneous starburst with solar metallicity, a Salpeter IMF, a lower mass cutoff of 
0.1\msunns, and upper mass cutoffs of 125\msun in the former and 100\msun in the 
latter. Although similar in gross behavior, the Starburst99 model predicts consistently 
redder (B$-$I) colors during the early phase of the burst as a result of nebular 
emission. The tick marks indicate the age of the starburst in Log(years).}

\figcaption{Comparison between Starburst 99 models for 0.05, 1, and 2 solar 
metallicities (i.e., Z=0.001, 0.02, and 0.04). The near-solar metallicity models are 
very similar, while the low metallicity model has consistently bluer colors. The 
tick marks indicate the age of the starburst in Log(years). Note that this figure 
has a different scale than Figure 6.}

\figcaption{Observed ({\it (U$-$B, B$-$I}) colors for warm (closed circles) and 
cool (open circles) ULIG nuclear regions, derived from the observations in 
Table 3 and the photometry in Paper I and Surace (1998). The known QSOs I Zw 1, Mrk 
1014, and IRAS 07598+6508 have been excluded. Also shown are the LH99 
starburst model (solid line) and the BC95 starburst (dotted line)  and 
continuous star formation (dashed-dotted line) models. The tick marks 
indicate the age of the starburst in Log(years). To illustrate the effects 
of dust reddening also plotted are a reddening vector (solid arrow) 
corresponding to A$_{\rm v} =$ 1 magnitude of foreground dust extinction, as 
well as the effects of mixed stars and dust in units of total V-band optical depth 
$\tau = $0, 5, 10, and 15 magnitudes. The derived colors are generally consistent 
with a young ($<$100 Myr) starburst, regardless of the evolutionary or reddening 
model.}

\figcaption{Same as Figure 8, but for double and single nucleus ULIGs. Squares 
represent those ULIGs with double nuclei but which do not have separate data points 
due to the limted spatial resolution at U\p.}

\clearpage
\vspace*{4.0truein}
Figures 1 \& 2  are images and are available in JPEG format from this preprint
archive.

\begin{figure}
\figurenum{3}
\epsscale{0.88}
\plotone{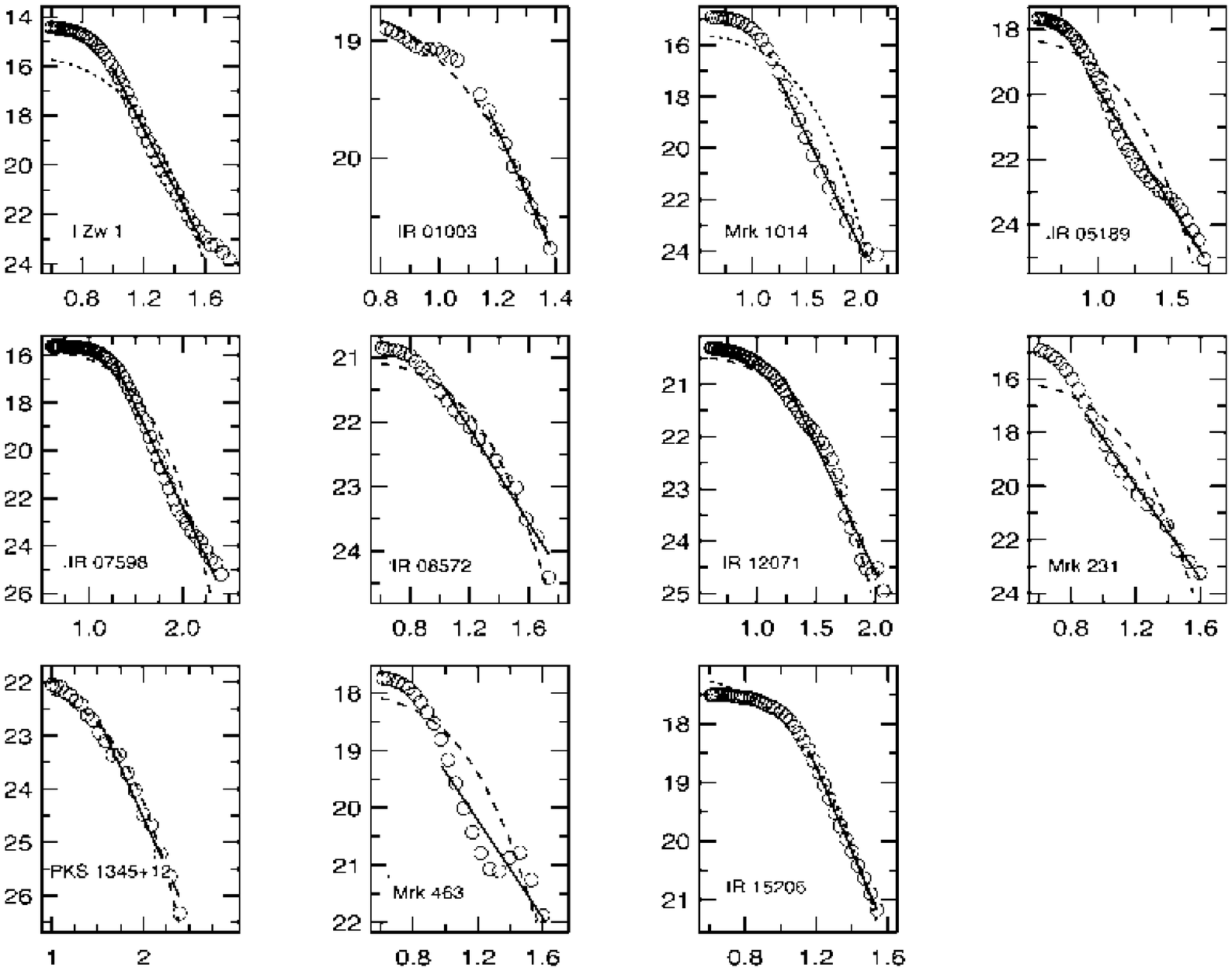}
\caption{}
\end{figure}
\clearpage

\begin{figure}
\figurenum{4}
\epsscale{0.9}
\plotone{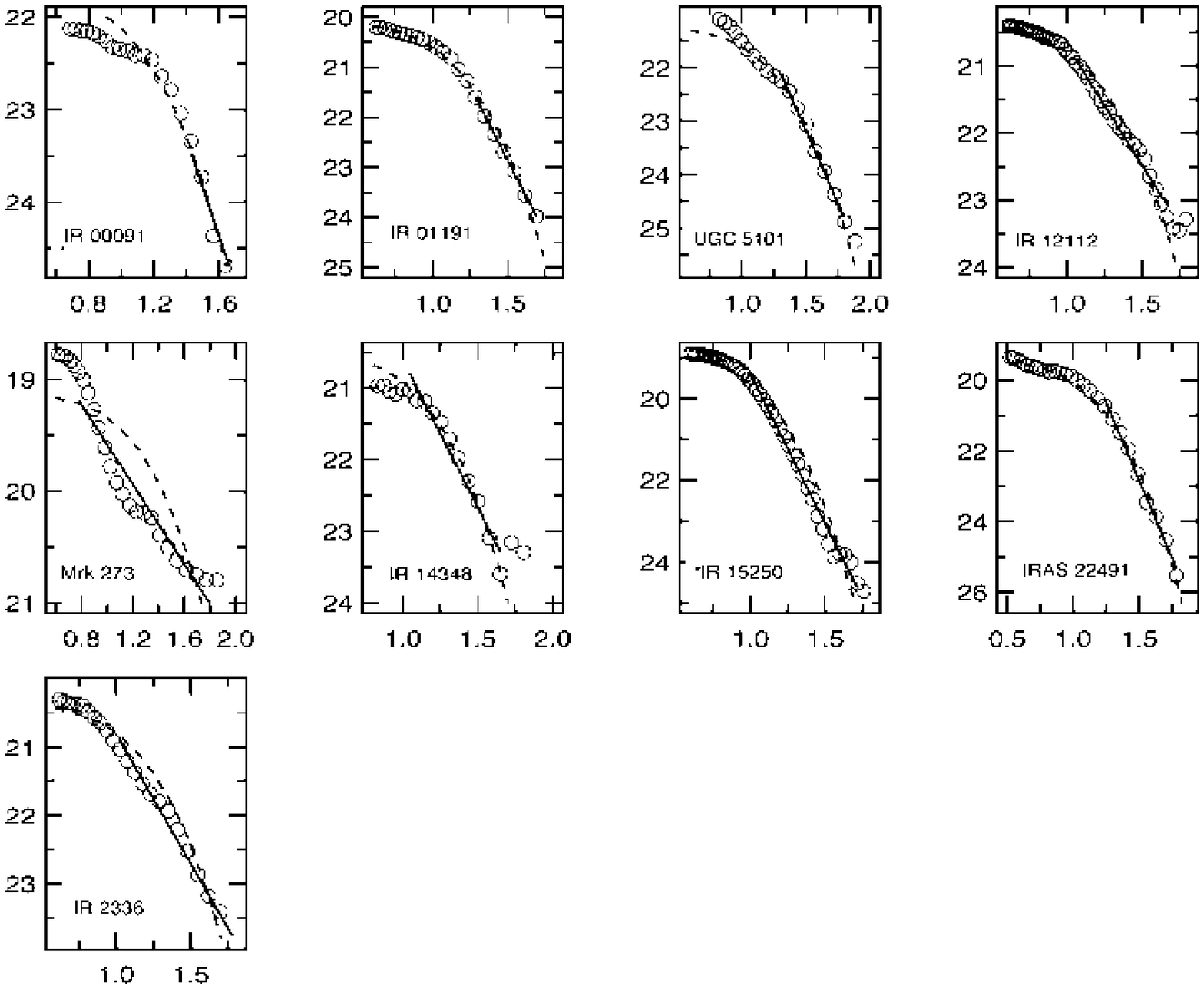}
\caption{}
\end{figure}
\clearpage

\begin{figure}
\figurenum{5}
\epsscale{0.9}
\plotone{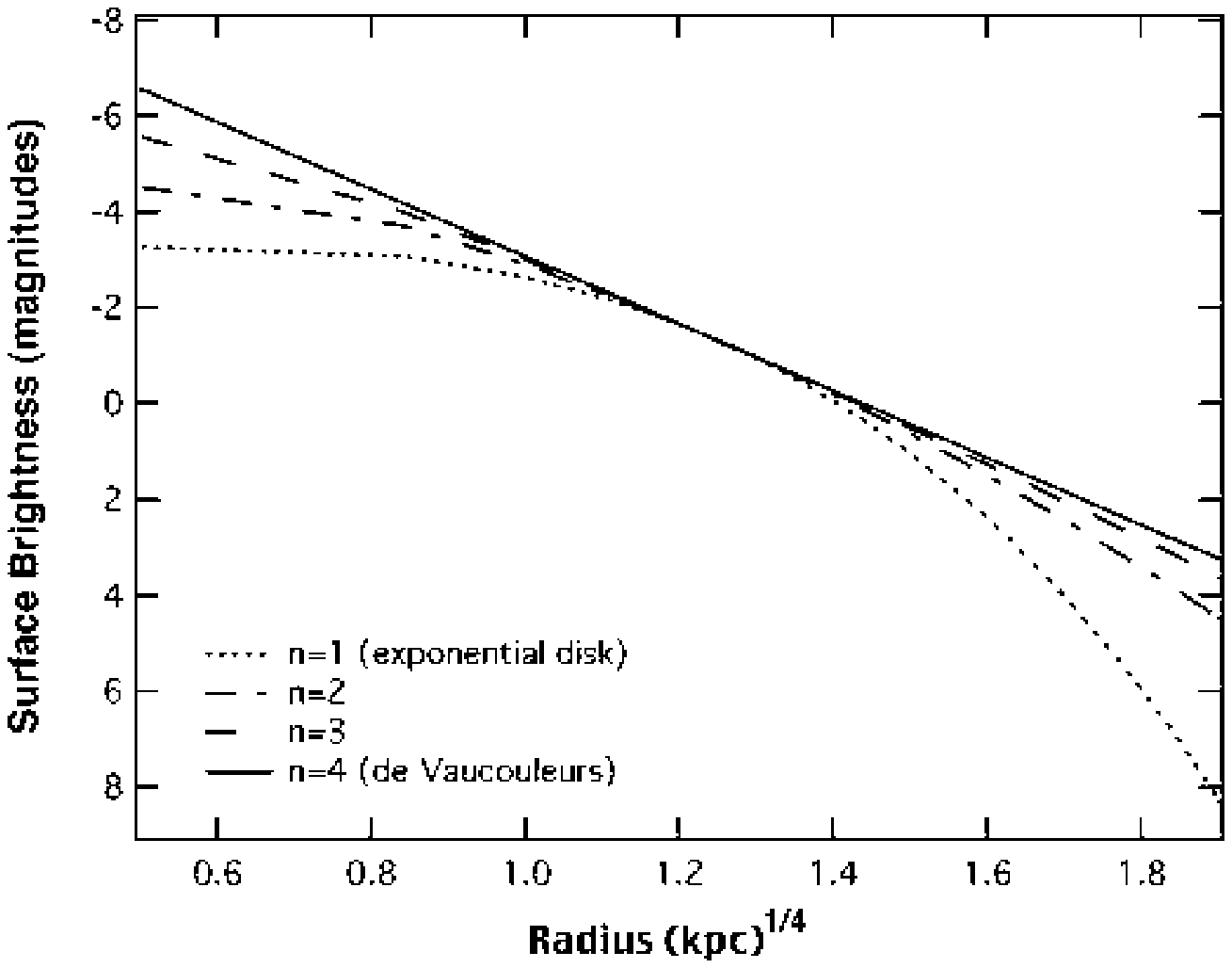}
\caption{}
\end{figure}
\clearpage

\begin{figure}
\figurenum{6}
\epsscale{0.9}
\plotone{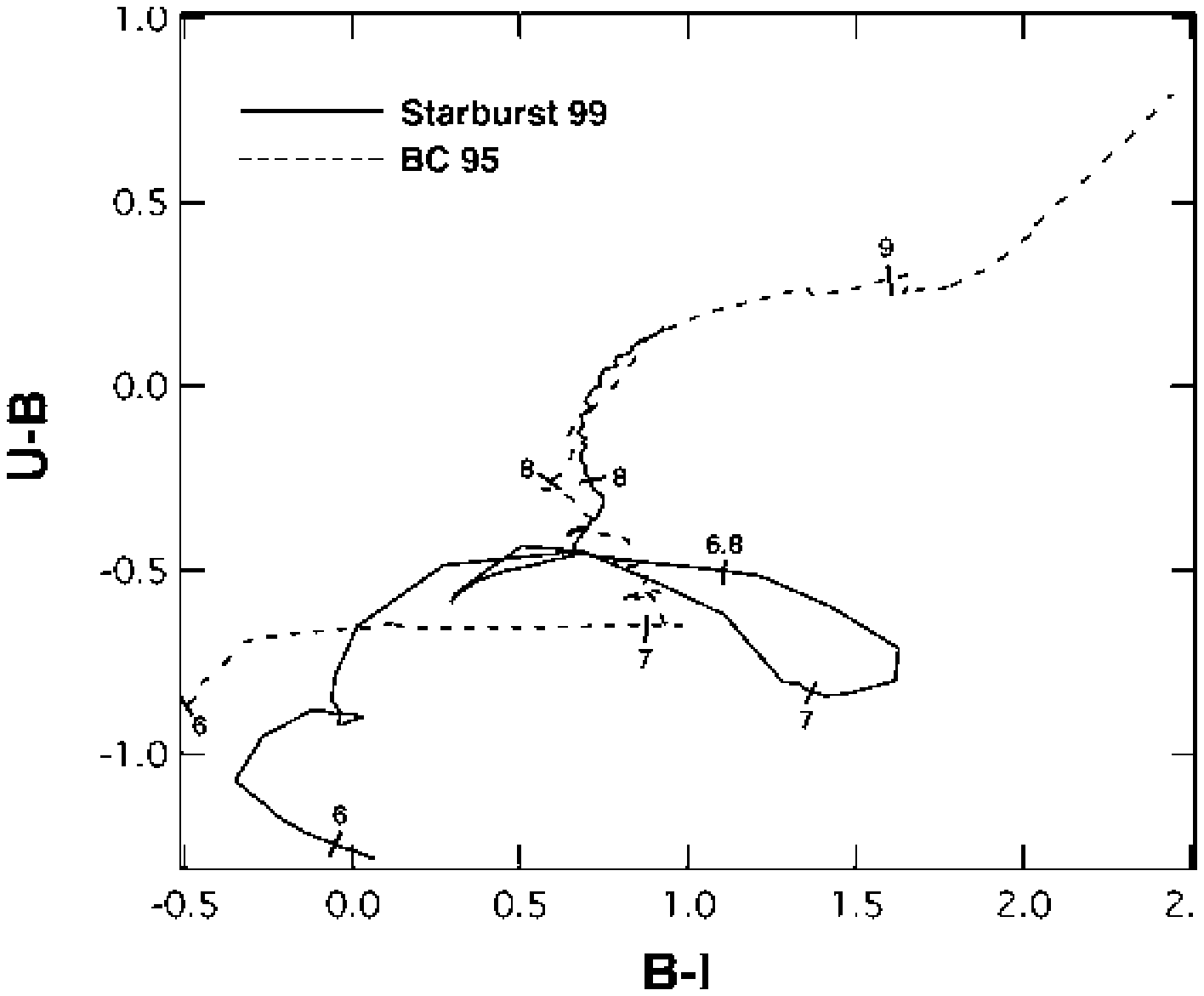}
\caption{}
\end{figure}
\clearpage

\begin{figure}
\figurenum{7}
\epsscale{0.9}
\plotone{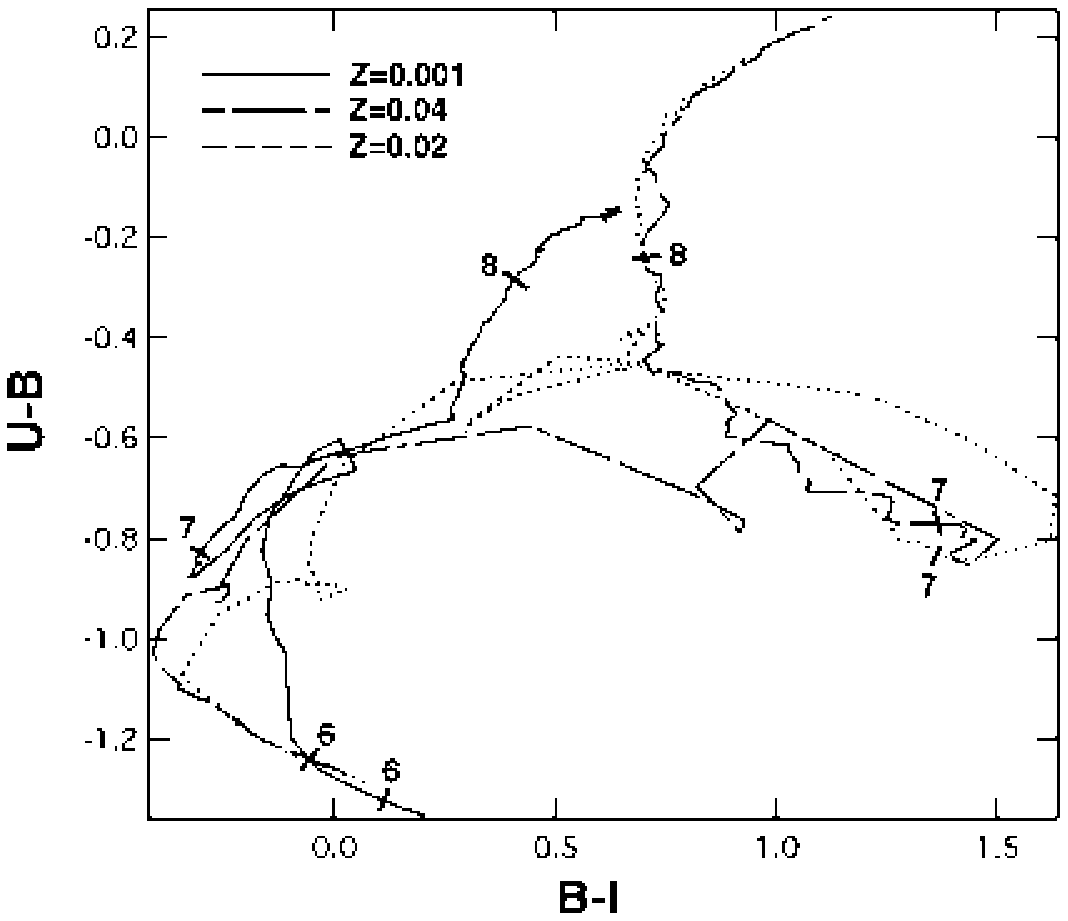}
\caption{}
\end{figure}
\clearpage

\begin{figure}
\figurenum{8}
\epsscale{0.9}
\plotone{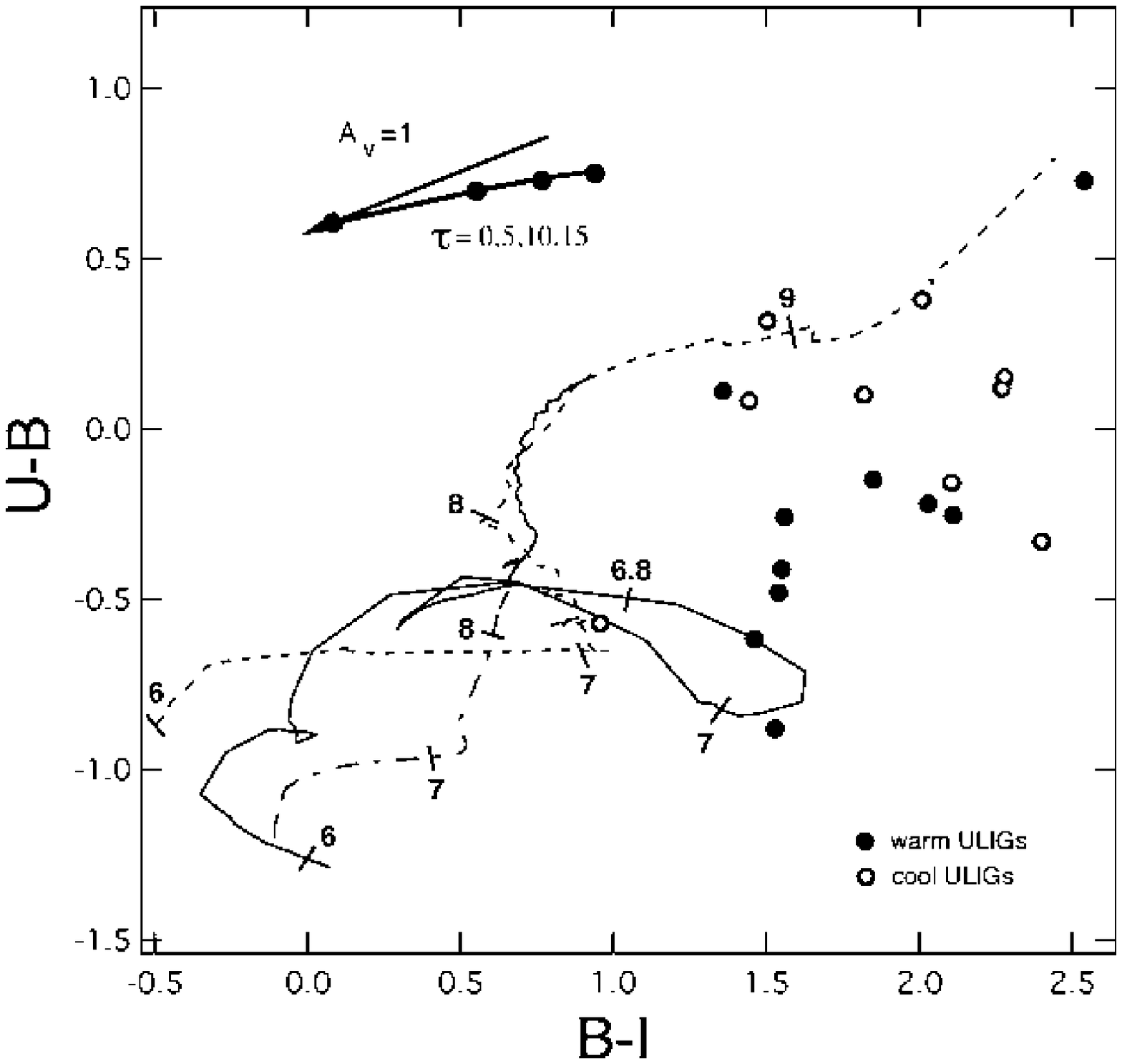}
\caption{}
\end{figure}
\clearpage

\begin{figure}
\figurenum{9}
\epsscale{0.9}
\plotone{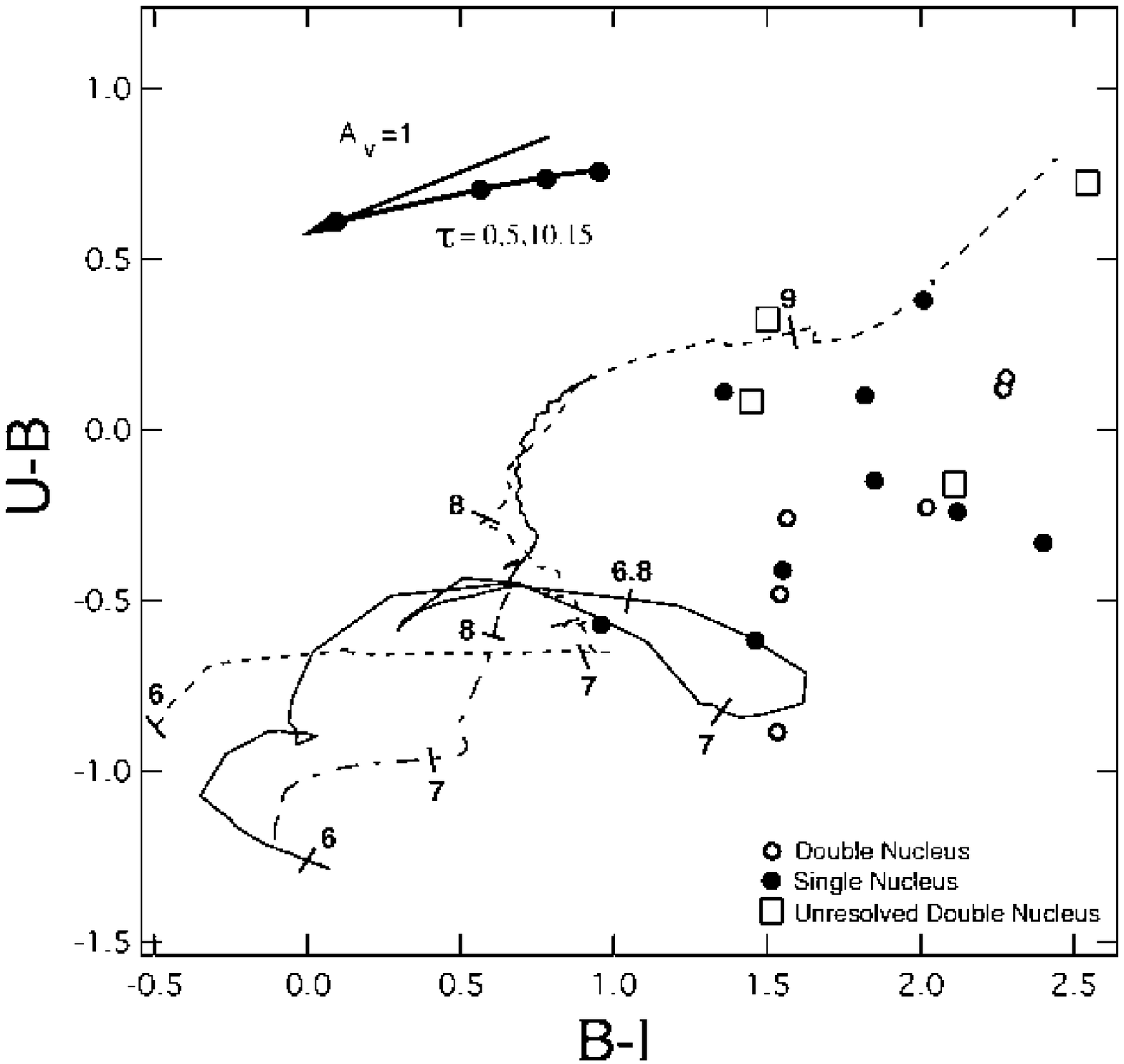}
\caption{}
\end{figure}
\clearpage

\clearpage

{\small
\begin{deluxetable}{lrrrcccc}
\tablewidth{0truein}
\tablecaption{Warm ULIG Sample}
\tablehead{
\colhead{Name} &
\colhead{RA} &
\colhead{DEC} &
\colhead{z} &
\colhead{Log {\it L}$_{\rm ir}$\tablenotemark{a}} &
\colhead{Exp. Time} &
\colhead{Nucleus} &
\colhead{Spectral Class} \\ [0.2ex]
\colhead{} &
\multicolumn{2}{c}{(J2000.0)} &
\colhead{} &
\colhead{{\it L}$_{\sun}$} &
\colhead{sec.}
}
\startdata
I{\ts}Zw{\ts}1          & 00:53:34.9 &  12:41:36.2 & 0.061 & 11.87 & 2400 & S & QSO \nl 
IRAS 01003$-$2238       & 01:02:49.8 & $-$22:21:56.3 & 0.118 & 12.24 & \phantom{1}360 & S & Wolf-Rayet \nl
Mrk 1014                & 01:59:50.1 &  00:23:41.5 & 0.163 & 12.53 & 3240 & S & QSO \nl
IRAS 05189$-$2524       & 05:21:01.5 & $-$25:21:46.7 & 0.042 & 12.07 & 6720 & S & Sy 2 \nl
IRAS 07598+6508		& 08:04:30.5 &  64:59:52.8 & 0.149 & 12.46 & 4000 & S & QSO \nl
IRAS 08572+3915         & 09:00:25.4 &  39:03:54.2 & 0.058 & 12.11 & 5040 & D & LINER \nl
IRAS 12071$-$0444       & 12:09:45.1 & $-$05:01:13.7 & 0.129 & 12.35& 3200 & S & Sy 2 \nl
Mrk 231                 & 12:56:14.2 &  56:52:26.1 & 0.042 & 12.50 & 1680 & S & QSO \nl
Pks 1345+12             & 13:47:33.5 &  12:17:24.5 & 0.122 & 12.28 & 3360 & D & Sy 2 \nl
Mrk 463                 & 13:56:02.7 &  18:22:18.3 & 0.051 & 11.77 & 3240 & D & Sy 1 \nl
IRAS 15206+3342         & 15:22:37.9 &  33:31:36.6 & 0.125 & 12.18 & 3480 & S & Sy 2 \nl
\enddata
\tablenotetext{a}{As given by Kim \& Sanders (1998), except I Zw 1 and Mrk 463 which are taken from Sanders et al. (1988b)}
\end{deluxetable}
}


{\small
\begin{deluxetable}{lrrrcccc}
\tablewidth{0truein}
\tablecaption{Cool ULIG Sample}
\tablehead{
\colhead{Name} &
\colhead{RA} &
\colhead{DEC} &
\colhead{z} &
\colhead{Log {\it L}$_{\rm ir}$\tablenotemark{a}} &
\colhead{Exp. Time} & 
\colhead{Nucleus\tablenotemark{b}} &
\colhead{Spectral Class} \\[.2ex]
\colhead{} &
\multicolumn{2}{c}{(J2000.0)} &
\colhead{} &
\colhead{{\it L}$_{\sun}$} &
\colhead{sec.}
}
\startdata
IRAS{\ts}00091$-$0738 & 00:11:43.2 & $-$07:22:07.8 & 0.119 & 12.19 & 2880 & S & HII \nl
IRAS{\ts}01199$-$2307 & 01:22:21.4 & $-$22:51:59.5 & 0.156 & 12.26 & 7200 & D & HII \nl
UGC{\ts}5101        & 09:35:51.7 &    61:21:11.3 & 0.039 & 12.01 & 4500 & S & Sy 1.5 \nl
IRAS{\ts}12112+0305 & 12:13:46.1 &    02:48:41.4 & 0.072 & 12.21 & 3240 & D & LINER \nl
Mrk{\ts}273         & 13:44:42.1 &    55:53:12.7 & 0.038 & 12.10 & 4000 & S & Sy 2 \nl
IRAS{\ts}14348$-$1447 & 14:37:38.7 & $-$15:00:22.8 & 0.082 & 12.28 & 3240 & D & LINER \nl
IRAS{\ts}15250+3609 & 15:26:59.4 &    35:58:37.5 & 0.055 & 12.00 & 2520 & S & LINER \nl
IRAS{\ts}22491$-$1808 & 22:51:49.3 & $-$17:52:23.5 & 0.078 & 12.08 & 3480 & D & LINER \nl
IRAS{\ts}23365+3604 & 23:39:01.3 &    36:21:09.8 & 0.064 & 12.10 & 4320 & S & LINER \nl
\enddata
\tablenotetext{a}{As given by Kim \& Sanders (1998), Kim (1995) and Sanders et al. (1988a)}
\tablenotetext{b}{S=single, D=double}
\end{deluxetable}
}


\tablenum{3}

{\small
\begin{deluxetable}{llcccr}
\tablewidth{5.75truein}
\tablecaption{U-band\tablenotemark{a}\ \  Observations of ULIGs}
\tablehead{
\colhead{Name} &
\colhead{ } &
\colhead{Aperture\tablenotemark{b}} &
\colhead{{\it m}$_{\rm U}$} &
\colhead{\it U$-$B} &
\colhead{Log $\nu${\it L}$_{\nu}$}  \\[.2ex]  
\multicolumn{2}{c}{\ } &
\colhead{\arcsec} &
\colhead{\ } &
\colhead{\ } &
\colhead{\hbox{L$_\odot$(3410\AA)}}
}
\startdata
IRAS{\ts}00091$-$0738 & Total & \nodata & 18.8 & \phantom{$-$}0.4 & 9.5 \nl
I{\ts}Zw{\ts}1 & Total & \nodata & 13.7 & $-$0.8 & 11.0 \nl
IRAS 01003$-$2238 & Total & \nodata & 17.9 & $-$0.6 & 9.9 \nl
IRAS 01199$-$2307 & Total & \nodata & 18.1 & \phantom{$-$}0.3 & 10.0 \nl
  & SW Nucleus & 3.5 & 18.9 & \phantom{$-$}0.5 & 9.7 \nl
  & NE Nucleus & 3.5 & 20.1 & \phantom{$-$}0.2 & 9.2 \nl 
Mrk{\ts}1014 & Total & \nodata & 14.7 & $-$1.2 & 11.4 \nl
 & Nucleus & 0.5 & 14.9 & $-$1.5 & 11.3 \nl
IRAS{\ts}05189$-$2524 & Total & \nodata & 15.6 & \phantom{$-$}0.0 & 9.9 \nl 
 & Nucleus & 1.5 & 16.2 & $-$0.2 & 9.6 \nl
IRAS 07598+6508 & Total & \nodata & 13.9 & $-$1.0 & 11.6 \nl
IRAS{\ts}08572+3915 & Total & \nodata & 16.7 & $-$0.1 & 9.7 \nl
 & SE Nucleus & 1.1 & 19.2 & $-$0.3 & 8.7 \nl
 & NW Nucleus & 1.1 & 19.6 & $-$0.2 & 8.6 \nl
 & 4 & 1.0 & 21.1 & $-$0.2 & 7.9 \nl
 & 5 & 1.0 & 21.2 & \phantom{$-$}0.0 & 7.9 \nl
UGC{\ts}5101 & Total & \nodata & 16.0 & \phantom{$-$}1.1 & 9.6 \nl
 & Nucleus & 1.6 & 19.1 & \phantom{$-$}1.2 & 8.2 \nl   
IRAS 12071$-$0444 & Total & \nodata & 17.7 & $-$0.5 & 10.0 \nl
IRAS 12112+0305 & Total & \nodata & 16.5 & $-$0.5 & 10.0 \nl
 & N Nucleus & 1.6 & 18.7 & $-$0.4 & 9.1 \nl
 & S Nucleus & 1.6 & $>$23.2\phantom{$<$} & $>$1.6 & $<$7.3 \nl
 & 1 & 1.6 & 19.3 & $-$0.2 & 8.9 \nl
 & 3 & 1.6 & 19.0 & $-$0.9 & 9.0 \nl
Mrk{\ts}231 & Total & \nodata & 14.1 & \phantom{$-$}0.3 & 10.5 \nl
 & Nucleus & 1.5 & 14.8 & \phantom{$-$}0.3 & 10.2 \nl 
 & 10/11/12/14/irA & \nodata & 17.9 & \phantom{$-$}0.2 & 8.9 \nl
 & 5/6 & \nodata & 18.8 & \phantom{$-$}0.1 & 8.6 \nl
Mrk{\ts}273 & Total & \nodata & 15.1 & \phantom{$-$}0.4 & 10.0  \nl
 & Nucleus & 1.7 & 17.2 & \phantom{$-$}0.1 & 9.2  \nl
 & S knot & 1.7 & 17.9 & \phantom{$-$}0.3 & 8.8 \nl
 & W knot & 1.7 & 17.8 & \phantom{$-$}0.4 & 8.9 \nl
Pks 1345+12 & Total & \nodata & 17.8 & \phantom{$-$}0.8 & 9.9 \nl
Mrk{\ts}463 & Total & \nodata & 14.8 & $-$0.6 & 10.4 \nl
 & E Nucleus & 1.3 & 16.5 & $-$0.9 & 9.7 \nl
 & W Nucleus & 1.3 & 16.8 & $-$0.5 & 9.5 \nl
 & 1/2/3/4 & 1.3 & 19.1 & $-$0.5 & 8.7\nl
 & 5/6 & 1.3 & 19.8 & $-$0.6  & 8.4 \nl
 & 7 & 1.3 & 20.6 & $-$0.3 & 8.1 \nl
 & 8/9/10 & 1.3 & 20.8 & $-$0.6 & 8.0 \nl 
IRAS 14348$-$1447 & Total & \nodata & 16.9 & \phantom{$-$}0.4 & 9.9 \nl
 & NE Nucleus & 3.0 & 18.7 & \phantom{$-$}0.1 & 9.2 \nl
 & SW Nucleus & 3.0 & 18.0 & \phantom{$-$}0.2 & 9.5 \nl
IRAS 15206+3342 & Total & \nodata & 16.2 & $-$0.6 & 10.6 \nl
 & E half & 1.3 & 16.5 &$-$0.9 & 10.4 \nl
 & W half & 1.3 & 18.1 & $-$0.4 & 9.8 \nl
IRAS 15250+3609 & Total & \nodata & 16.2 & \phantom{$-$}0.0 & 9.9 \nl
 & Nucleus & 1.2 & 17.0 & $-$0.6 & 9.6 \nl
IRAS 22491-1808 & Total & \nodata & 17.2 & \phantom{$-$}0.7 & 9.8 \nl
 & W. Nucleus,2,3 & 1.5 & 18.1 & \phantom{$-$}0.1 & 9.4 \nl
IRAS{\ts}23365+3604 & Total & \nodata& 16.8 & \phantom{$-$}0.7 & 9.8 \nl
 & Nucleus & 1.0 & 19.2 & $-$0.3 & 8.8 \nl
 & 1 & 1.0 & 19.8 & $-$0.1 & 8.6 \nl
 & 2 & 0.5 & 20.7 & $-$0.8 & 8.2 \nl 
\enddata
\tablenotetext{a}{The actual observations were made in the U$^{\prime}$ filter, and then converted via equation 1 into the Johnson U-band magnitudes
reported here.}
\tablenotetext{b}{Aperture radii are given where applicable. ``Nucleus'' values are in most cases for a circular aperture 2.5 kpc in diameter with appropriate aperture corrections.} 
\tablecomments{Knot identification numbers refer to those in Surace et al. (1998) and Surace (1998),
and should be referenced as SSVVM(1998):Number or Surace(1998):Number.}
\end{deluxetable}
}


\tablenum{4}
{\small
\begin{deluxetable}{lcccr}
\tablewidth{5.75truein}
\tablecaption{Morphological Features\tablenotemark{a}}
\tablehead{
\colhead{Name} &
\colhead{mid-IR\tablenotemark{b}} &
\colhead{Single/Double Nucleus} &
\colhead{Tails\tablenotemark{c}} &
\colhead{Knots in Tails} \\ [2.0ex]  
}
\startdata
IRAS 01199$-$2307 & C & D & Y & N \nl
IRAS 08572+3915 & W & D & Y & Y \nl
IRAS 14348$-$1447 & C & D & Y & Y \nl
Pks 1345+12 & W & D & Y & N \nl
Mrk 463 & W & D & Y & Y \nl
IRAS 12112+0305 & C & D & Y & Y \nl
IRAS 22491$-$1808 & C & D & Y & Y \nl
IRAS 05189$-$2524 & W & S & Y & N \nl
Mrk 1014 & W & S & Y & Y \nl
IRAS 15206+3342 & W & S & N & N \nl
Mrk 231 & W & S & Y & N \nl
IRAS 00091-0738 & C & S & Y & N \nl
IRAS 23365+3604 & C & S & Y & N \nl
UGC 5101 & C & S & Y & N \nl
Mrk 273 & C & S & Y & N \nl
IRAS 15250+3609 & C & S & Y & N \nl
IRAS 12071-0444 & W & S & Y & N \nl
IRAS 01003-2238 & W & S & N & N \nl
I Zw 1 & W & S & Y & Y \nl
IRAS 07598+6508 & W & S & N & N \nl
\enddata
\tablenotetext{a}{Ordering based on single/double nuclei and luminosity-normalized tail length as given in Surace (1998)}
\tablenotetext{b}{W=warm, C=cold, as defined in the text.} 
\tablenotetext{c}{As detected at optical/near-infrared wavelengths (Surace et al. 1998,2000). Knots are as detected at U$^{\prime}$ in the extended tail regions seen at I-band (Surace et al. 1999,2000).}
\end{deluxetable}
}


\clearpage

\begin{references}
\reference{allen} Allen, C.W. {\it Astrophysical Quantities}, 1973, (London: Athlone 
Press)
\reference{} Bahcall, J.N., Kirhakos, S., Saxe, D.H., \& Schneider, D.P. 1997, \apj, 
479, 642
\reference{amy} Barger, A.J., Cowie, L.L., Sanders, D.B., Fulton, E., Taniguchi, Y., 
Sato, Y., Kawara, K., \& Okuda, H., 1998, \nat, 394, 248
\reference{} Bessell, M.S. 1979, \pasp, 91, 589
\reference{} Bruzual, G., \& Charlot, S. 1993, \apj, 405, 538  
\reference{} Bryant, P., \& Scoville, N.Z., 1996, \apj, 457, 678
\reference{} Clements, D.L., Sutherland, W.J., McMahon, R.G., \& Saunders, W. 
1996, \mnras, 279, 477
\reference{} Charlot, S., Ferrari, F., Mathews, G.J., \& Silk, J. 1993, \apj, L57
\reference{} de Grijp, M.H.K., Lub, J., \& Miley, G., 1987, \aap, 70, 95
\reference{} de Grijp, M.H.K., Miley, G., Lub, J., \& de Jong, T. 1985, \nat, 314, 
240
\reference{} Guiderdoni, B., \& Rocca-Volmerange, B., 1988, \apss, 74, 185
\reference{} Elvis, M. 1994, \apjs, 95, 1
\reference{} Elvis, M., Green, R.F., Bechtold, J., Schmidt, M., et al. 1986, \apj, 
310, 291
\reference{fanelli} Fanelli, M. N. , Waller, W. W. , Smith, D. A. , Freedman, W. L. , Madore, B.  , 
Neff, S. G. , O'Connell, R. W. , Roberts, M. S. , Bohlin, R.  , Smith, A. M. , Stecher, T. P., 
1997, \apj, 481, 735
\reference{hill} Hill, J. K. , Cheng, K. -P. , Bohlin, R. C. , Cornett, R. H. , Hintzen, P. M. N. , 
O'Connell, R. W. , Roberts, M. S. , Smith, A. M. , Smith, E. P. , Stecher, T. P. 1995, \apj, 438, 
181
\reference{hogg} Hogg, D.W., Pahre, M.A., McCarthy, J.K., Cohen, J.G., Blandford, R., 
Smail, I., \& Soifer, B.T., 1997, \mnras, 288,404
\reference{} Kim, D.-C. 1995, Ph.D. Thesis, University of Hawaii
\reference{} Kim, D-C., \& Sanders, D.B., 1998, \apj, 119, 41
\reference{knapen} Knapen, J.H., Laine, S., Yates, J.A., Robinson, A., Richards, A., 
Doyon, R., \& Nadeua, D. 1997, \apj, 490, L29
\reference{landolt83} Landolt, A., 1983, \aj, 88, 439
\reference{landolt92} Landolt, A., 1992, \aj, 104, 340
\reference{larson} Larson, R.B., \& Tinsley, B.M., 1978, \apj, 219, 46
\reference{leitherer} Leitherer, C. 1996, in From Stars to Galaxies, eds. C. 
Leitherer, U. Fritze-vonAlvensleben, J. Huchra (San Francisco: ASP), 373
\reference{} Leitherer, L., \& Heckman, T. 1995, \apjs, 96, 9
\reference{} Leitherer, C., Schraerer, D., Goldader, J.D., Delgado, , R.M., Robert, 
C., Kune, D.F., et al. 1998, in press (Starburst99)
\reference{low} Low, F.J., Cutri, R.M., Huchra, J.P., \& Kleinmann, S., 1988, \apjl, 
327, L41
\reference{meurer} Meurer, G., Heckman, T., Leitherer, C., Kinney, A., Robert, 
C., \& Garnett, D. 1995, \apj, 110, 2665
\reference{murphy} Murphy, T., Armus, L., Matthews, K., Soifer, B.T., 
Mazzarella, J.M., Shupe, D.L., Strauss, M.A., \& Neugebauer, G. 1996, \aj, 111, 
1025
\reference{} Neugebauer, G., Green, R.F., Matthews, K., Schmidt, M., Soifer, B.T., 
\& Bennett, J. 1987, \apjs, 63, 615
\reference{peebles} Peebles, P.J., 1993, Principles of Physical Cosmology, (Princeton 
University Press: Princeton), 327
\reference{perault} Perault, M. 1987, Ph.D. Thesis, University of Paris
\reference{} Press, W.H., Teukolsky, S.A., Vettering, W.T., \& Flannery, B.P., 1992, 
``Numerical Recipes in C'', (Cambridge University Press: Cambridge)
\reference{rieke} Rieke, G.H., \& Lebofsky, M.J. 1985, \apj, 288, 619
\reference{} Rowan-Robinson, M.; Mann, R.G.; Oliver, S. J.; Efstahiou, A.; Eaton, 
N.; Goldschidt, P.; Mobasher, B.; Serjeant, S. B. G.; Sumner, T.  .et al. 1997, 
\mnras, 289, 490
 \reference{sanders-araa} Sanders, D.B., \& Mirabel, I.F. 1996, \araa, 34, 749
\reference{sanders-bgs} Sanders, D.B., Soifer, B.T., Elias, J.H., Madore, B.F., 
Matthews, K., Neugebauer, G., \& Scoville, N.Z. 1988a, \apj, 325, 74
\reference{} Sanders, D.B., Soifer, B.T., Elias, J.H., Neugebauer, G., \& Matthews, 
K., 1988b, \apjl, 328, L35
\reference{} Schweizer, F., 1982, \apj, 252, 455
\reference{scoville} Scoville, N.Z., Yun, M.S., \& Bryant, P.M., 1997, \apj, 484, 702
\reference{scoville2} Scoville, N.Z., Evans, A.S., Thompson, R., Rieke, M., Hines, D., Low, 
F.J., Dinshaw, N., Surace, J.A., \& Armus, L., 1999, \aj, in prep
\reference{} Sersic, J.L., 1968, Atlas de Galaxias Australis (Cordoba: 
Observatorio Astronomica)
\reference{} Smail, I., Ivison,R.J., \& Blain, A.W., 1997, \apjl, 490, L5
\reference{} Smail, I., Ivison,R.J., Blain, A.W., \& Kneib, J.-P., 1998, \apjl, 507, 
L21
\reference{smith} Smith, D. A. , Neff, S. G. , Bothun, G. D. , Fanelli, M. N. , Offenberg, J. D. 
, Waller, W. H. , Bohlin, R. C. , O'Connell, R. W. , Roberts, M. S. , Smith, A. M. , \& Stecher, 
T. P. 1996, \apjl, 473, L21
\reference{soifer} Soifer, B.T., Neugebauer, G., Matthews, K., Egami, E., Becklin, E.E., 
Weinberger, A.J., Ressler, M., Werner, M.W., Evans, A.S., Scoville, N.Z., Surace, J.A., \& 
Condon, J.J, 2000, \aj, in press
\reference{stanford} Stanford, S.A., \& Bushouse, H., 1991, \apj, 371, 92
\reference {stockton} Stockton, A., \& MacKenty, J.W., 1987, \apj, 316, 584
\reference{} Surace, J.A. 1998, Ph.D. Thesis, University of Hawaii
\reference{} Surace, J.A., Sanders, D.B., Vacca, W.D., Veilleux, S., \& Mazzarella, 
J.M., 1998, \apj, 492, 116 (Paper I)
\reference{} Surace, J.A., \& Sanders, D.B., 1999, \apj, 512, 162,  (Paper II)
\reference{} Surace, J.A., Sanders, D.B., \& Evans, A.S. 2000, \apj, in press
\reference{toomre} Toomre, A., 1977, in ``The Evolution of Galaxies and Stellar 
Populations'', (Yale Observatories: New Haven)
\reference{neil} Trentham, N., Kormendy, J., \& Sanders, D.B., 1999, \aj, in press
\reference{wainscoat} Wainscoat, R.J., 1996, ``The UH 2.2m Telescope Reference 
Manual'', 
(University of Hawaii: Honolulu)
\reference{whit} Whitmore, B., \& Schwiezer, F. 1995, \aj, 109, 960 
\reference{wright} Wright, G.S., James, P.A., Joseph, R.D., \& McLean, I.S., 1990, \nat, 
344, 417
\reference{young} Young, S., Hough, J.H., Efstathiou, A., Wills, B.J., Bailey, J.A., Ward, 
M.J., \& Axon, D.J. 1996, \mnras, 281, 1206
\reference{zheng} Zheng, Z., Wu, H., Mao, S., Xia, X.-Y., Deng, Z.-G., \& Zhou, Z.-L., \aa, 
in press
\end{references}
\end{document}